\documentclass[preprint,nofootinbib]{revtex4}%
\usepackage{amssymb}
\usepackage{amsfonts}
\usepackage{amsmath}
\usepackage{graphicx}
\usepackage[usenames]{color}%
\begin{document}
\title{\begin{flushright}
{ \small CECS-PHY-08/05\\ \vspace{-.3cm}DAMTP-2008-38 }
\end{flushright}\vskip1.0cm Stability of asymptotically AdS wormholes in
vacuum against scalar field perturbations}
\author{Diego H. Correa$^{1}$, Julio Oliva$^{2,4}$, and Ricardo Troncoso$^{2,3}$}
\affiliation{$^{1}$DAMTP, Centre for Mathematical Sciences, University of Cambridge,
Wilberforce Road, Cambridge CB3 0WA, UK.}
\affiliation{$^{2}$Centro de Estudios Cient\'{\i}ficos (CECS), Casilla 1469, Valdivia, Chile.}
\affiliation{$^{3}$Centro de Ingenier\'{\i}a de la Innovaci\'{o}n del CECS (CIN), Valdivia, Chile.}
\affiliation{$^{4}$Departamento de F\'{\i}sica, Universidad de Concepci\'{o}n, Casilla,
160-C, Concepci\'{o}n, Chile.}
\email{D.Correa@damtp.cam.ac.uk, juliooliva@cecs.cl, ratron@cecs.cl}

\begin{abstract}
The stability of certain class of asymptotically AdS wormholes in
vacuum against scalar field perturbations is analyzed. For a free
massive scalar field, the stability of the perturbation is
guaranteed provided the squared mass is bounded from below by a
negative quantity. Depending on the base manifold of the AdS
asymptotics, this lower bound could be more stringent than the
Breitenlohner-Freedman bound. An exact expression for the spectrum
is found analytically. For a scalar field perturbation with a
nonminimal coupling, slow fall-off asymptotic behavior is also
allowed, provided the squared mass fulfills certain negative upper
bound. Although the Ricci scalar is not constant, an exact
expression for the spectrum of the scalar field can also be found,
and three different quantizations for the scalar field can be
carried out. They are characterized by the fall-off of the scalar
field, which can be fast or slow with respect to each asymptotic
region. For these perturbations, stability can be achieved in a
range of negative squared masses which depends on the base manifold
of the AdS asymptotics. This analysis also extends to a
class of gravitational solitons with a single conformal boundary.
\end{abstract}
\maketitle
\tableofcontents

\section{Introduction}

The existence of wormhole solutions, describing handles in the spacetime
topology, is an interesting question that has been raised repeatedly in
theoretical physics within different subjects, and it is as old as General
Relativity. The systematic study of this kind of objects in the static case,
was pushed forward by the seminal works of Morris, Thorne and Yurtsever
\cite{MTY}, which have shown that requiring the existence of exotic matter
that violates the standard energy conditions around the throat is inevitable
(For a review see \cite{Visser}). Because of that, the stability as well as
the existence of wormholes remains somehow controversial. Exotic matter is
also required to construct static wormholes for General Relativity in higher
dimensions. Nonetheless, in higher-dimensional spacetimes, if one follows the
same basic principles of General Relativity to describe gravity, the Einstein
theory is not the only possibility. Indeed, the most general theory of gravity
in higher dimensions leading to second order field equations for the metric is
described by the Lovelock action which possesses nonlinear terms in the
curvature \cite{LOVELOCK}. Within this framework, it is worth pointing out
that in five dimensions it has been found that the so-called
Einstein-Gauss-Bonnet theory, being quadratic in the curvature, admits static
wormhole solutions in vacuum \cite{DOTWORM}. Precisely, these solutions were
found allowing a cosmological (volume) term in the Einstein-Gauss-Bonnet
action, and choosing the coupling constant of the quadratic term such that the
theory admits a single anti-de Sitter (AdS) vacuum. These wormholes connect
two asymptotically locally AdS spacetimes each with a geometry at the boundary
that is not spherically symmetric. These solutions extend to higher odd
dimensions for special cases of the Lovelock class of theories, also selected
by demanding the existence of a unique AdS vacuum. Generically, the mass of
the wormhole appears to be positive for observers located at one side of the
neck, and negative for the ones at the other side, such that the total mass
always vanishes. This provides a concrete example of \textit{mass without
mass}. The apparent mass at each side of the wormhole vanishes only when the
solution acquires reflection symmetry. In this case the metric reads
\begin{equation}
ds_{d}^{2}=l^{2}\left(  -\cosh^{2}\rho\ dt^{2}+d\rho^{2}+\cosh^{2}%
\rho\ d\Sigma_{d-2}^{2}\right)  \,,\label{metricworm}%
\end{equation}
where $d\Sigma_{d-2}$ stands for the line element of a $\left(
d-2\right) $-dimensional base manifold $\Sigma_{d-2}$. The metric
(\ref{metricworm}) is an exact solution of the aforementioned
special class of gravity theories in odd dimensions $d=2n+1$ greater
than three, provided $\Sigma_{2n-1}$ satisfies Eq. (\ref{TrCS})
presented in appendix B. It is worth to remark that no energy
conditions are violated by the solution (\ref{metricworm}) since the
whole spacetime is devoid of any kind of stress-energy tensor. Thus,
it is natural to wonder whether this wormhole can be regarded as a
stable solution providing a suitable ground state to define a field
theory on it.

As a first step in this direction, here we study the stability of scalar field
perturbations on the wormholes described by the metric (\ref{metricworm}).
These perturbations are stable provided their squared masses satisfy a lower
bound, which is generically more restrictive than that discovered by
Breitenlohner and Freedman for AdS spacetime \cite{BF1}, given by
$m^{2}>m_{BF}^{2}$ with
\begin{equation}
m_{BF}^{2}:=-\frac{1}{l^{2}}\left(\frac{d-1}{2}\right)^{2}\,.
\label{mbeefe}%
\end{equation}

The metric (\ref{metricworm}) describes a wormhole with a neck
located at $\rho=0$, connecting two asymptotically AdS spacetimes
but with a more general compact base manifold $\Sigma_{d-2}$ without
boundary. Explicit examples of base manifolds that solve equation
(\ref{TrCS}) are presented in appendix B, where we also present the
solutions for base manifolds that are all the possible products of
constant curvature spaces in five and seven dimensions. In all these
examples the base manifolds possess locally hyperbolic
factors $H_n$ which can be quotiented by a freely acting
discrete subgroup of $O(n,1)$ such that the quotient
becomes smooth and compact. The solutions with non-quotiented
hyperbolic factors are in fact not wormholes, but instead describe
gravitational solitons with a single conformal boundary. The
analysis of the stability of scalar field perturbations performed
here also extends for these gravitational solitons when
they are endowed with an end of the world brane.

\bigskip

In this paper, we consider the stability of scalar perturbations on
wormhole geometries of the form (\ref{metricworm}) with an arbitrary
base manifold in any dimension, thus including the solutions for
class of theories mentioned above. The strategy we follow is similar
to the one used by Breitenlohner and Freedman for AdS$_4$ spacetime
\cite{BF1}, \cite{BF2} (and by Mezincescu and Townsend in their generalization to
$d$ dimensions \cite{MEZTOW}). In those original works, the allowed scalar field fluctuations
on AdS (and their asymptotic fall-off) are determined either by looking at the energy functional
and demanding it to converge or by looking at the energy flux at the spatial infinity and
demanding it to vanish. Following those criteria, fluctuations with slow fall-off were allowed in
the range of masses $m^2_{BF} < m^2 < m^2_{BF} + \tfrac{1}{l^2}$, when the  stress-energy tensor
was improved and for a precise value of the improvement coefficient. In particular, we will adopt the
criterion of the vanishing energy flux at the spatial infinity. Since the statement about the allowed
fluctuations depends only on the AdS asymptotics of the spacetime, we should observe that the slow
fall-off fluctuations on (\ref{metricworm}) are admissible in the same range of masses and for the
same value of the non-minimal coupling as in AdS spacetime.

In the next section, we solve the Klein-Gordon equation for a free massive scalar field minimally
coupled to gravity. Remarkably, this can be done analytically on the
background metric (\ref{metricworm}), so that an exact expression
for the spectrum is found requiring the energy flux to vanish at
each boundary. These boundary conditions single out the fast
fall-off asymptotic behavior\footnote{The asymptotic radial behavior
of the scalar field at the boundaries will be typically of the form
$e^{-2|\rho|\lambda _{\pm}}$ and we refer to branches
$\lambda_{\pm}$ as fast and slow fall-off respectively (see below).}
for the scalar fields. Consequently, it is shown that stability of
(\ref{metricworm}) under these free massive scalar perturbations is
guaranteed provided the squared mass is bounded from below by a
negative quantity which depends on the lowest eigenvalue of the
Laplace operator on the base manifold.

In Section III it is shown that, in the presence of nonminimal coupling with
the scalar curvature, scalar fields with slow fall-off also give rise to
conserved energy perturbations, which are stable provided the negative squared
mass also satisfies certain upper bound. It is worth to remark that, unlike
the case of AdS spacetime, the Ricci scalar of the wormhole is not constant,
so that the nonminimal coupling contributes to the field equation with more
than a mere shift in the mass. Nevertheless, in this case an exact expression
for the spectrum can also be found, and three different quantizations for the
scalar field can be carried out, being characterized by the fall-off of the
scalar field, which can be either fast or slow in each one of the asymptotic
regions. The range of masses where these perturbations are stable will depend
on the base manifold of the AdS asymptotics. Section IV is devoted to final
comments and remarks. The asymptotic expansions of the generalized Legendre
functions describing the radial fall-off of the scalar field on the wormhole
are presented in the appendix A. Appendix B is devoted to concrete examples
capturing the features described above, where a thorough analysis of the
seven-dimensional case is performed for base manifolds that are all the
possible products of constant curvature spaces.

\section{Exact spectrum of free massive scalar fields}

Let us consider the line element (\ref{metricworm}). As explained above, this
metric describes a wormhole with a neck located at $\rho=0$. Remarkably, this
background metric allows to find an analytic expression for the spectrum of a
free massive scalar field $\phi$ satisfying the Klein-Gordon equation
\begin{equation}
\left(  \square-m^{2}\right)  \phi=0\,. \label{KG}%
\end{equation}
This can be seen adopting the following ansatz
\begin{equation}
\phi=e^{-i\omega t}f(\rho)Y(\Sigma)\,, \label{separation}%
\end{equation}
where $Y\left(  \Sigma\right)  $ is an eigenfunction of the Laplace
operator on the base manifold $\Sigma_{d-2}$, i.e.\footnote{If the
base manifold is assumed to be compact and without boundary, then
$Q\geq0$.}, $\nabla^{2}Y=-QY$. Hence, the radial function $f(\rho)$
has to satisfy the following equation:
\begin{equation}
\frac{d^{2}f(\rho)}{d\rho^{2}}+\left(  d-1\right)  \tanh\rho\frac{df(\rho
)}{d\rho}+\left(  \frac{\omega^{2}-Q}{\cosh^{2}\rho}-m^{2}l^{2}\right)
f(\rho)=0\,. \label{radialfecu}%
\end{equation}
It is convenient to change the coordinates as
\begin{equation}
x=\tanh\rho\,,
\end{equation}
so that the boundaries at $\rho\rightarrow\pm\infty$ are now located at
$x\rightarrow\pm1$. It is also useful to express the radial function as
\begin{equation}
f(x)=\left(  1-x^{2}\right)  ^{\frac{d-1}{4}}K(x)\,,
\end{equation}
such that (\ref{radialfecu}) reduces to a Legendre equation for $K(x)$, i.e.
\begin{equation}
(1-x^{2})\frac{d^{2}K(x)}{dx^{2}}-2x\frac{dK(x)}{dx}+\left(  \nu(\nu
+1)-\frac{\mu^{2}}{1-x^{2}}\right)  K(x)=0\,, \label{legendre}%
\end{equation}
where the parameters $\mu$ and $\nu$ are defined by
\begin{align}
\mu &  :=\sqrt{\left(  \frac{d-1}{2}\right)  ^{2}+m^{2}l^{2}}\,,\\
\nu &  :=\sqrt{\left(  \frac{d-2}{2}\right)  ^{2}+\omega^{2}-Q}-\frac{1}{2}\,.
\end{align}
The general solution of Eq. (\ref{legendre}) is given by an arbitrary linear
combination of the associated Legendre functions of first and second kind,
$P_{\nu}^{\mu}(x)$ and $Q_{\nu}^{\mu}(x)$, respectively. If $\mu$ is not an
integer\footnote{The case in which $\mu$ is an integer is discussed separately
at the end of this section.}, the solution is conveniently expressed as an
arbitrary linear combination of $P_{\nu}^{\mu}(x)$ and $P_{\nu}^{-\mu}(x)$.
Therefore, the general solution of the radial equation (\ref{radialfecu}) is
given by
\begin{equation}
f(x)=\left(  1-x^{2}\right)  ^{\frac{d-1}{4}}\left[  C_{1}P_{\nu}^{\mu
}(x)+C_{2}P_{\nu}^{-\mu}(x)\right]  \,, \label{efeoriginal}%
\end{equation}
where $C_{1}$ and $C_{2}$ are integration constants. Thus, according to the
asymptotic behavior of the Legendre functions (see Appendix \ref{A}), the
radial function $f(x)$ admits, at each boundary, two possible asymptotic
behaviors with leading terms $(1\pm x)^{\lambda_{+}}$ and $(1\pm
x)^{\lambda_{-}}$, with
\begin{equation}
\lambda_{\pm}:=\frac{d-1}{4}\pm\frac{\mu}{2}\,. \label{lambdas}%
\end{equation}
The asymptotic form of $f(x)$ near the boundaries is then given by
\begin{equation}
f(x)\underset{x\rightarrow\mp1}{\sim}\!C_{1}\left(  1\pm x\right)
^{\lambda_{-}}\left[  \alpha_{\pm}+\mathcal{O}\left(  1\pm x\right)  \right]
+C_{2}\left(  1\pm x\right)  ^{\lambda_{+}}\left[  \beta_{\pm}+\mathcal{O}%
\left(  1\pm x\right)  \right]  \,,
\end{equation}
where $\alpha_{\pm}$ and $\beta_{\pm}$ are constants. In analogy with the case
of AdS spacetime, for $m^{2}>0$ the $\lambda_{-}$ branch leads to a divergent
behavior of the scalar field at the boundaries, so that only the $\lambda_{+}$
branch is admissible. Nonetheless, for $m_{BF}^{2}<m^{2}<0$, where $m_{BF}%
^{2}$ is defined in (\ref{mbeefe}), both branches $\lambda_{+}$ and
$\lambda_{-}$ are allowed in principle, corresponding to fast and slow
fall-off respectively.

Then, the asymptotic behavior of the scalar field is determined by
imposing suitable boundary conditions. In order to ensure the
conservation of the energy of the scalar field, we require the
vanishing of the energy flux at both spatial infinities
$x\rightarrow\pm1$. For wormhole geometries, it is natural to
consider both conditions separately since two boundaries exist.

Note that for the cases in which the metric (\ref{metricworm})
describes a gravitational soliton there is only one conformal
boundary, since the non-compact hyperbolic factors of the base
manifold are joined at the boundary, i.e., the boundaries of the
corresponding Poincar\'{e} balls are identified at spatial
infinities, $x\rightarrow\pm1$. Nevertheless, wormhole-like boundary
conditions for $x\rightarrow\pm1$ can also be extended to this case
provided the solitons are endowed with an end of the world brane.
This is the analogue of the  ``non-transparent" boundary conditions
considered in AdS when an end of the world brane is located at
$\rho=0$ \cite{KR, Porrati, MM}\footnote{It could be very
interesting to analyze the possibility of introducing such a brane
completely in vacuum in the same lines of Refs. \cite{HTZBrane, GW,
GGGW}. However this is beyond the scope of this work. }. Hereafter,
for the sake of completeness we assume that the gravitational
solitons are always endowed with an end of the world brane, so that
the analysis of their stability against scalar field perturbations
with ``non-transparent" boundary conditions can be
straightforwardly borrowed from the one of wormholes.

The energy current is given by $j^{\mu}=\sqrt{-g}\eta^{\nu}T_{\text{ \ }\nu
}^{\mu}$, where $\eta$ is the time-like Killing vector $\partial_{t}$ and
$T_{\mu\nu}$ the stress-energy tensor for the free massive scalar field,
\begin{equation}
T_{\mu\nu}=\partial_{\mu}\phi\partial_{\nu}\phi-\frac{1}{2}g_{\mu\nu}%
g^{\alpha\beta}\partial_{\alpha}\phi\partial_{\beta}\phi-\frac{m^{2}}{2}%
g_{\mu\nu}\phi^{2}\,.
\end{equation}
Hence, the radial energy flux goes like
\begin{equation}
\sqrt{-g}g^{\rho\rho}T_{\rho0}\sim(1-x^{2})^{-\frac{d-3}{2}}\partial
_{x}\left(  \phi^{2}\right)  \,, \label{rflux}%
\end{equation}
and its vanishing at each boundary reduces to
\begin{equation}
\left.  (1-x^{2})^{-\frac{d-3}{2}}f(x)\frac{df(x)}{dx}\right\vert _{x=\pm
1}=0\,. \label{flux}%
\end{equation}
These two boundary conditions will determine a discrete spectrum of
frequencies for the scalar field perturbation.

Using the general solution (\ref{efeoriginal}) to evaluate Eq. (\ref{flux}) at
$x\rightarrow+1$, one obtains
\begin{align}
&  (1-x)^{-\mu}\left[  A_{0}(\mu)^{2}C_{1}^{2}(d-1-2\mu)+\mathcal{O}%
(1-x)\right] \nonumber\\
&  +\left[  2A_{0}(\mu)A_{0}(-\mu)C_{1}C_{2}(d-1)+\mathcal{O}(1-x)\right]
\label{flux1}\\
&  +(1-x)^{\mu}\left[  A_{0}(-\mu)^{2}C_{2}^{2}(d-1+2\mu)+\mathcal{O}%
(1-x)\right]  =0\,.\nonumber
\end{align}
The constant $A_{0}$ in this asymptotic behavior comes directly from the
asymptotic expansion of $P_{\nu}^{\mu}(x)$ (see Eq. (\ref{As1})). Thus, the
vanishing of the energy flux at $x\rightarrow+1$ requires the vanishing of the
first and second line of (\ref{flux1}). The only possibility is $C_{1}%
=0$\footnote{The vanishing of the factor $d-1-2\mu$ is not a possibility. In
odd dimensions it would correspond to an integer value of $\mu$, while in even
dimensions the coefficients of the sub-leading terms in the first line of
(\ref{flux1}) would be still non-vanishing. Requiring $A_{0}(\mu)=0$ would
also imply that $\mu$ is an integer.}, so that only the fast fall-off behavior
$f(x)\sim\left(  1-x\right)  ^{\lambda_{+}}$ is allowed for $x\rightarrow+1$.

Analogously, evaluating (\ref{flux}) at the other boundary located at
$x\rightarrow-1$, one obtains
\begin{align}
&  (1+x)^{-\mu}\left[  D_{0}(-\mu)^{2}(d-1-2\mu)+\mathcal{O}(1+x)\right]
\nonumber\\
&  +\left[  2D_{0}(-\mu)B_{0}(-\mu)(d-1)+\mathcal{O}(1+x)\right]
\label{flux-1}\\
&  +(1+x)^{\mu}\left[  B_{0}(-\mu)^{2}(d-1+2\mu)+\mathcal{O}(1+x)\right]
=0\,,\nonumber
\end{align}
where the constants $B_{0}\ $and $D_{0}$ are defined in Eqs. (\ref{B0}) and
(\ref{D0}), respectively.

Note that all the non-vanishing terms in the limit $x\rightarrow-1$ of Eq.
(\ref{flux-1}) possess a factor $1/\Gamma(\mu-\nu)$ (see Eqs. (\ref{D0}) and
(\ref{D1})). Thus, the flux at $x=-1$ also vanishes, provided the coefficients
satisfy $\mu-\nu=-n$, with $n$ a non-negative integer. This restriction
determines a discrete spectrum of frequencies and also singles out the
$\left(  1-x\right)  ^{\lambda_{+}}$ behavior in the asymptotic region
$x\rightarrow-1$ (fast fall-off).

It remains to consider the case when $\mu=k$ is integer. Then, we use $P_{\nu
}^{k}(x)$ and $Q_{\nu}^{k}(x)$ as the linearly independent solutions of
(\ref{legendre}),
\begin{equation}
f\left(  x\right)  =\left(  1-x^{2}\right)  ^{\frac{d-1}{4}}\left[
C_{1}P_{\nu}^{k}\left(  x\right)  +C_{2}Q_{\nu}^{k}\left(  x\right)  \right]
\, \label{solmuentero}%
\end{equation}
The solution $Q_{\nu}^{k}(x)$ always contributes to the flux with logarithmic
divergencies at infinity. Thus, turning off $Q_{\nu}^{k}(x)$ the energy flux
at $x\rightarrow+1$ vanishes\footnote{For $k=0$, the energy flux is
non-vanishing even for $C_{2}=0.$}. The vanishing of the radial flux in the
other asymptotic region is accomplished by demanding $k-\nu=-n$ as for the
generic case.

\bigskip

Therefore, for real $\mu$, the relation $\mu-\nu=-n$ gives the spectrum of
frequencies, which reads
\begin{equation}
\omega^{2}=\left(  n+\frac{1}{2}+\sqrt{\left(  \frac{d-1}{2}\right)
^{2}+m^{2}l^{2}}\right)  ^{2}-\left(  \frac{d-2}{2}\right)  ^{2}+Q\,.
\label{rrfrec}%
\end{equation}

Let us now analyze the stability of these scalar perturbations. For
non-negative squared masses, stability is guaranteed independently of the base
manifold, since in this case $\omega^{2}>0$.

Furthermore, as it occurs for AdS spacetime, perturbations with $m^{2}<0$ are
also allowed. To what extent it is possible to take negative values for
$m^{2}$ depends on the base manifold. For a generic base manifold
$\Sigma_{d-2}$, if $\left(  \frac{d-2}{2}\right)  ^{2}-Q-\frac{1}{4}$ is
always non-positive, $\omega^{2}$ given in (\ref{rrfrec}) will be always
positive and the only restriction on the mass comes from the BF bound, i.e.,
$m^{2}>m_{BF}^{2}$. To decide if the positivity of $\omega^{2}$ imposes any
condition on the mass, it suffices to consider the lowest mode ($n=0$) and the
lowest eigenvalue of the Laplace operator on $\Sigma_{d-2}$, denoted by
$Q_{0}$. It turns out that whenever $Q_{0}<\left(  \frac{d-2}{2}\right)
^{2}-\frac{1}{4}$, stability ($\omega^{2} > 0$) imposes a lower bound to the
squared mass being more stringent than that of Breitenlohner-Freedman.

The general case can be summarized with the following bound
\begin{equation}
m^{2}>m_{BF}^{2}+m_{\Sigma}^{2}\,, \label{imporvedbf}%
\end{equation}
with $m_{BF}^{2}$ defined in (\ref{mbeefe}) and
\begin{equation}
m_{\Sigma}^{2}l^{2}:=\left\{
\begin{array}
[c]{cc}%
\left[  \sqrt{\left(  \frac{d-2}{2}\right)  ^{2}-Q_{0}}-\frac{1}{2}\right]
^{2} & :\ Q_{0}<\left(  \frac{d-2}{2}\right)  ^{2}-\frac{1}{4}\\
0 & :\ Q_{0}\geq\left(  \frac{d-2}{2}\right)  ^{2}-\frac{1}{4}%
\end{array}
\right.  \,. \label{mimprovement}%
\end{equation}
Therefore, stability is guaranteed provided the bound (\ref{imporvedbf}) with
(\ref{mimprovement}) is fulfilled.

\bigskip

In order to visualize the dependence of the bound (\ref{imporvedbf})
on the base manifold $\Sigma_{d-2}$, it is useful to consider some
specific examples. This is performed here for maximally symmetric
spaces in diverse dimensions, even beyond the ones for which
(\ref{metricworm}) is a solution in vacuum for the special class of
odd-dimensional gravity theories. Nevertheless, we will stress the
cases of base manifolds constituting vacuum solutions.

\bigskip

$\circ$ If the base manifold $\Sigma_{d-2}$ corresponds to a torus
$T^{d-2}$, or a sphere $S^{d-2}$, then the lowest eigenvalue of the
Laplace operator is $Q_{0}=0$. Hence, the squared frequencies
(\ref{rrfrec}) remain positive as long as the mass is bounded by a
negative quantity
\begin{equation}
m^{2}l^{2}>-\left(  d-2\right)  \,, \label{Bound-Q0=0-Fast}%
\end{equation}
which nonetheless, is more stringent than the BF bound.

\bigskip

$\circ$ If $\Sigma_{d-2}$ is a manifold of negative constant
curvature, it must be the hyperbolic space $H_{d-2}$ or a smooth
quotient thereof. A case of special interest is when $H_{d-2}$ is of
unit radius, because the metric (\ref{metricworm}) is a vacuum
solution of a special class of higher-dimensional gravity theories
\cite{DOTWORM} (see appendix B).  For non-quotiented $H_{d-2}$, the
spacetime describes a gravitational soliton. In this case, the
spectrum of the Laplace operator takes the form
\begin{equation}
Q=\left(  \frac{d-3}{2}\right)  ^{2}+\zeta^{2}\,,\label{QH}%
\end{equation}
where the parameter $\zeta$ takes all real values. As the lowest eigenvalue of
the Laplace operator is $Q_{0}=\left(  \frac{d-3}{2}\right)  ^{2}$, the
squared frequencies remain positive provided
\begin{equation}
m^{2}l^{2}>-\frac{d^{2}-4d+5+2\sqrt{2d-5}}{4}\,,\label{Hfullmin}%
\end{equation}
which is also more stringent than the BF bound.

Upon quotients such that $H_{d-2}/\Gamma$ is a closed surface of
finite volume, the metric (\ref{metricworm}) corresponds to a
wormhole. The spectrum of the Laplace operator  becomes a discrete
set and the zero mode, $Q=0$, should also be included, so that the
bound is given by Eq.(\ref{Bound-Q0=0-Fast}).

\bigskip

$\circ$ As a last example, we consider $\Sigma_{d-2}=S^{1}\times
H_{d-3}$, where the radius of $H_{d-3}$ is fixed to $\left(
d-2\right)  ^{-1/2}$. With this choice, the  metric
(\ref{metricworm}) is also a vacuum solution of the mentioned
higher-dimensional gravity theory \cite{DOTWORM}  (see appendix B).
 For a gravitational soliton with non-quotiented $H_{d-3}$, the lowest eigenvalue of
 the Laplace operator is $Q_0=(d-2)\left(\tfrac{d-4}{2}\right)^2$, which always satisfies
 $Q_{0}<\left(\tfrac{d-2}{2}\right)^{2}-\tfrac{1}{4}$ except for $d=5$. Hence stability
 of this soliton is guaranteed provided
\begin{equation}
m^{2}l^{2}>\left\{
\begin{array}
[c]{cc}%
-\frac{9+2\sqrt{6}}{4} & :\ d=5\\
m_{BF}^{2}l^{2} & :\ d>5
\end{array}
\right.  \,, \label{S1Hd-3min}%
\end{equation}
where the bound is more stringent than that of
Breitenlohner-Freedman only in five dimensions. Again, considering a
quotient of $H_{d-3}$, such that the metric (\ref{metricworm})
describes a wormhole solution, stability is achieved for the bound
(\ref{Bound-Q0=0-Fast}).

\bigskip

In sum, in this section it has been shown that scalar field
perturbations on the metric (\ref{metricworm}) are stable provided
that the squared mass satisfies a negative lower bound given by
(\ref{mimprovement}). Depending on the lowest eigenvalue of the
Laplace operator on the base manifold, this bound can be more
stringent than the BF bound.  When (\ref{metricworm}) is a wormhole
solution this bound is always (\ref{Bound-Q0=0-Fast}).

So far, we have considered minimally coupled free scalar field
perturbations on the metric (\ref{metricworm}). Although the
Klein-Gordon equation admits two different behaviors at the
asymptotic regions, after imposing the vanishing of the energy flux
at the spatial infinities, only the fast fall-off at both boundaries
is allowed. As it is shown in the next section, for certain ranges
of negative squared mass, one can also satisfy the boundary
conditions with slow fall-off scalar fields by \textquotedblleft
improving\textquotedblright\ the stress-energy tensor with a term
coming from a non-minimal coupling of the scalar field with the
scalar curvature of the background geometry.

\section{Scalar fields with nonminimal coupling}

\label{nonminimal coupling}

It is well-known that in AdS spacetime, improving the stress-energy tensor
with a term coming from a non-minimal coupling of the scalar field with
gravity, allows to include the slow fall-off branch within the
spectrum\footnote{For locally AdS spacetimes describing massless topological
black holes with hyperbolic base manifolds \cite{Mann1,Mann2,Vanzo, BLP,
Birmingham, Cai-Soh, Aros}, scalar fields with slow fall-off are also allowed
\cite{QNM}, provided the mass of the scalar field satisfies the bound
$m_{BF}^{2}<m^{2}<m_{BF}^{2}+l^{-2}$. This also guarantees its stability under
gravitational perturbations, since they reduce to scalar field perturbations
with different masses corresponding to scalar, vector and tensor modes
\cite{BM1, BM2}. Its perturbative stability under gravitational perturbations
has also been analyzed in \cite{Gibbons-Hartnoll, Kodama-Ishibashi, Kodama}.
The nonperturbative stability can be ensured from the fact that they admit
Killing spinors for certain class of base manifolds \cite{Groundstate}.}
\cite{BF1}. Remarkably, an exact expression for the spectrum of a scalar field
coupled to the non-constant Ricci scalar is also found for the three different
quantizations that can be carried out, depending of the fall-off of the scalar
field at each asymptotic region.

\bigskip

Let us now consider a scalar field perturbation on the wormhole geometry
(\ref{metricworm}) including a nonminimal coupling with the scalar
curvature\footnote{The conformal coupling is recovered for $\xi=\frac{1}%
{4}\frac{d-2}{d-1}$.  The propagation of conformally coupled scalar
fields on asymptotically AdS backgrounds has been studied in
\cite{Mann3}.},
\begin{equation}
\left(  \square-m^{2}-\xi R\right)  \phi=0\,, \label{KGNM}%
\end{equation}
where $R$ is the Ricci scalar of the background metric (\ref{metricworm}),
given by
\begin{equation}
R=-\frac{d\left(  d-1\right)  }{l^{2}}~+\frac{(d-1)\left(  d-2\right)
+\tilde{R}}{l^{2}\cosh^{2}\left(  \rho\right)  }\,. \label{erre}%
\end{equation}
Here $\tilde{R}$ is the Ricci scalar of the base manifold $\Sigma_{d-2}$,
which is assumed to be constant in order to ensure the separability of the
wave equation (\ref{KGNM}). Note that, unlike the case of AdS spacetime, the
Ricci scalar of the wormhole given by (\ref{erre}) is not constant, so that
the nonminimal coupling contributes now to the field equation (\ref{KGNM})
with more than a mere shift in the mass. Indeed, the effect of the additional
contribution given by the second term at the r.h.s. of (\ref{erre}) amounts to
a shift in the frequency term in Eq.(\ref{radialfecu}), so that the total
effect of the nonminimal coupling will entail corrections containing $\xi$ in
both parameters $\mu$ and $\nu$.

Performing separation of variables as in Eq. (\ref{separation}), the equation
for the radial function reduces to
\begin{equation}
\frac{d^{2}f\left(  \rho\right)  }{d\rho^{2}}+\left(  d-1\right)  \tanh
\rho\frac{df\left(  \rho\right)  }{d\rho}+\left(  \frac{\omega
_{\text{\textit{eff}}}^{2}-Q}{\cosh^{2}\rho}-m_{\text{\textit{eff}}}^{2}%
l^{2}\right)  f\left(  \rho\right)  =0\,. \label{ecuradialeff}%
\end{equation}
It is worth pointing out that one obtains the same equation as in the case of
minimal coupling, which has already been solved in the previous section, but
with an effective mass and frequency given by
\begin{align}
\omega_{\text{\textit{eff}}}^{2}  &  :=\omega^{2}-\xi\left[  \left(
d-1\right)  \left(  d-2\right)  +\tilde{R}\right]  \,,\label{fef}\\
m_{\text{\textit{eff}}}^{2}l^{2}  &  :=m^{2}l^{2}-d\left(  d-1\right)  \xi\,.
\label{mef}%
\end{align}
Hence, the solution of (\ref{ecuradialeff}) can be written as in
(\ref{efeoriginal}) if $\mu$ is not an integer, and it is given by Eq.
(\ref{solmuentero}) for $\mu=k$, with $k$ an integer, where now
\begin{align}
\nu &  =\sqrt{\left(  \frac{d-2}{2}\right)  ^{2}+\omega_{\text{\textit{eff}}%
}^{2}-Q}-\frac{1}{2}\,,\\
\mu &  =\sqrt{\left(  \frac{d-1}{2}\right)  ^{2}+m_{\text{\textit{eff}}}%
^{2}l^{2}}\,, \label{mueff}%
\end{align}
are defined in terms of the effective frequency and mass, given by (\ref{fef})
and (\ref{mef}), respectively.

As explained in the previous section, the general solution for the scalar
field possesses two possible asymptotic fall-offs at each boundary. The
presence of a nonminimal coupling affects the vanishing of the energy flux
boundary condition, in such a way that it can be compatible with slow fall-off
scalar fields.

Let us see how the nonminimal coupling modifies the boundary conditions. The
stress-energy tensor for the nonminimally coupled scalar field acquires the
form
\begin{equation}
T_{\mu\nu}=\partial_{\mu}\phi\partial_{\nu}\phi-\frac{1}{2}g_{\mu\nu}%
g^{\alpha\beta}\partial_{\alpha}\phi\partial_{\beta}\phi-\frac{m^{2}}{2}%
g_{\mu\nu}\phi^{2}+\xi\left[  g_{\mu\nu}\square-\nabla_{\mu}\nabla_{\nu
}+G_{\mu\nu}\right]  \phi^{2}\,,
\end{equation}
so that, requiring the energy flux to vanish at both infinities, one obtains
\begin{equation}
\left(  1-x^{2}\right)  ^{-\frac{\left(  d-1\right)  }{2}}\left[  \left(
1-4\xi\right)  \left(  1-x^{2}\right)  \frac{df(x)}{dx}f(x)+2\xi
\ xf(x)^{2}\right]  _{x=\pm1}=0\,. \label{flux2}%
\end{equation}
Using the asymptotic expansion (\ref{As1}), the condition (\ref{flux2}) at
$x\rightarrow+1$ reduces to
\begin{align}
&  (1-x)^{-\mu}\left(  -A_{0}(\mu)^{2}C_{1}^{2}\left(  1+\left(
2\mu-d\right)  \left(  1-4\xi\right)  \right)  +\mathcal{O}[1-x]\right)
\nonumber\\
&  +\left(  2A_{0}(\mu)A_{0}(-\mu)C_{1}C_{2}((1-4\xi)(d-1)-4\xi)+\mathcal{O}%
[1-x]\right) \label{flux3}\\
&  +(1-x)^{\mu}\left(  -A_{0}(-\mu)^{2}C_{2}^{2}\left(  1+\left(
2\mu-d\right)  \left(  1-4\xi\right)  \right)  +\mathcal{O}[1-x]\right)
=0\,.\nonumber
\end{align}

If $C_{1}=0$, then (\ref{flux3}) automatically vanishes, and one obtains the
fast fall-off at $x\rightarrow+1$. Nevertheless, the presence of a nonminimal
coupling, allows switching on the branch with slow fall-off, since the first
line in (\ref{flux3}) can also vanish for $C_{1}\neq0$. This can be done by
choosing $\xi$ such that
\begin{equation}
(1+(2\mu-d)(1-4\xi))=0\, ,
\end{equation}
and $\mu<1$, i.e.,
\begin{equation}
\xi=\xi_{0}:=\frac{\lambda_{-}}{1+4\lambda_{-}} \,. \label{beta}%
\end{equation}
In order to ensure the vanishing of the second line of (\ref{flux3}), it is
necessary to impose $C_{2}=0$, which singles out the slow fall-off of the
scalar field at $x\rightarrow+1$. Note that for the branch with slow fall-off,
the condition $\mu<1$ imposes a negative upper bound on the effective squared
mass, given by
\begin{equation}
m^{2}<m_{BF}^{2}+\frac{1}{l^{2}}\,. \label{mbef+1}%
\end{equation}
Notice that the range of masses $m^2_{BF} < m^2 < m^2_{BF} + \tfrac{1}{l^2}$
as well as the specific value of the nonminimal coupling (\ref{beta}) are exactly the same as the ones
allowing slow fall-off scalar fluctuations on AdS spacetime \cite{BF1}, \cite{BF2}, \cite{MEZTOW}.

Let us now analyze condition (\ref{flux2}) at $x\rightarrow-1$. When one
chooses the fast fall-off at $x\rightarrow+1$, with $C_{1}=0$, the asymptotic
expansion of (\ref{flux2}) for $x\rightarrow-1$ reduces to
\begin{align}
&  \!\!\! (1+x)^{-\mu}\left(  \left(  -1\right)  ^{d}C_{2}^{2}D_{0}(-\mu
,\nu)^{2}\left(  1+\left(  2\mu-d\right)  \left(  1-4\xi\right)  \right)
+\mathcal{O}[1+x]\right) \nonumber\\
&  \!\!\! +\left(  -2\left(  -1\right)  ^{d}C_{2}^{2}D_{0}(-\mu,\nu)B_{0}%
(-\mu,\nu)(1-d+4\xi d)+\mathcal{O}[1+x]\right) \label{flux4}\\
&  \!\!\! +(1+x)^{\mu}\left(  \left(  -1\right)  ^{d}C_{2}^{2}B_{0}(-\mu
,\nu)^{2}\left(  \left(  2\mu+d\right)  \left(  1-4\xi\right)  -1\right)
+\mathcal{O}[1+x]\right)  =0\, .\nonumber
\end{align}

In order to fulfill Eq. (\ref{flux4}), one possibility is to impose
$D_{0}(-\mu,\nu)=0$, where $D_{0}\left(  \mu,\nu\right)  $ is defined in
(\ref{D0}). This singles out the fast fall-off at $x\rightarrow-1$, and
implies that $\mu-\nu=-n$ with $n$ a nonnegative integer. This quantization
relation gives the spectrum corresponding to fast fall-off at both sides of
the wormhole, hereafter referred as \textit{fast-fast fall-off}.

The other possibility is to require the vanishing of $B_{0}(-\mu,\nu)$, with
$\mu<1$, and $\xi=\xi_{0}$, where $\xi_{0}$ is given by (\ref{beta}). In this
case, the branch with slow fall-off is selected at $x\rightarrow-1$. As
defined in (\ref{B0}), $B_{0}(-\mu,\nu)$ vanishes for $\nu=n$, with $n$ a non
negative integer. This condition gives the spectrum corresponding to the
\textit{fast-slow fall-off}. As shown below, this spectrum differs from the
one obtained previously for fast-fast fall-off.

As explained above, the slow fall-off at $x\rightarrow+1$ is singled out by
imposing simultaneously in (\ref{flux3}) $C_{2}=0$, $\xi=\xi_{0}$ and $\mu<1$.
In this case, the condition (\ref{flux2}) at $x\rightarrow-1$ reduces to
\begin{align}
&  \!\!\!\! \left(  2\left(  -1\right)  ^{d}C_{1}^{2}D_{0}(\mu,\nu)B_{0}%
(\mu,\nu)(d\left(  1-4\xi_{0}\right)  -1)+\mathcal{O}[1+x]\right) \nonumber\\
&  \!\!\!\! +(1+x)^{\mu}\left(  -\left(  -1\right)  ^{d}C_{1}^{2}D_{0}(\mu
,\nu)^{2}\left(  \left(  2\mu+d\right)  \left(  1-4\xi_{0}\right)  -1\right)
+\mathcal{O}[1+x]\right)  =0\, , \label{flux5}%
\end{align}
so that (\ref{flux5}) can vanish by requiring either $B_{0}(\mu,\nu)=0$ or
$D_{0}(\mu,\nu)=0$. The former condition corresponds to the fast fall-off at
$x\rightarrow-1$. In this case, the quantization condition again reads $\nu=n$
with $n$ a nonnegative integer. This is naturally expected, since this case
corresponds to the slow-fast fall-off, which is obtained from the case with
fast-slow fall-off, by the reflection symmetry of the wormhole metric
(\ref{metricworm}) with respect to $\rho=0$.

Finally, the condition $D_{0}(\mu,\nu)=0$ implies $\mu+\nu=n$ or
$1-\mu +\nu=-n$ where $n$ is a nonnegative integer. Both conditions
conduce to the same spectrum, and this case corresponds to the
\textit{slow-slow fall-off}.

In a similar fashion, cases $\mu=k$ with $k$ integer are shown to admit the
same type of spectra.

\bigskip

So far, we have shown that at each boundary, the scalar field presents two
possible behaviors, one corresponding to fast fall-off with a leading term
that behaves as $(1\pm x)^{\frac{d-1}{4}+\frac{\mu}{2}}$, and the other
corresponding to the slow fall-off whose leading term behaves as $(1\pm
x)^{\frac{d-1}{4}-\frac{\mu}{2}}$. Let us now consider the spectra coming from
the three possible quantizations and analyze the stability of these
nonminimally coupled excitations:

\begin{itemize}
\item Fast-fast fall-off
\end{itemize}

In this case the scalar field possesses fast fall-off at both sides of the
wormhole and the spectrum is obtained from the quantization relation
\begin{equation}
\mu-\nu=-n\,,
\end{equation}
so that the frequencies are given by
\begin{equation}
\omega^{2}\!=\!\left(  n+\frac{1}{2}+\sqrt{\left(  \frac{d-1}{2}\right)
^{2}\!+m_{\text{\textit{eff}}}^{2}l^{2}}\right)  ^{2}\!\!-\left(  \frac
{d-2}{2}\right)  ^{2}\!\!+Q+\xi\left[  \left(  d-1\right)  \left(  d-2\right)
+\tilde{R}\right]  \,. \label{rrfrec2}%
\end{equation}

Let us recall that in this case the value of the coupling constant $\xi$ is
not restricted. If the following condition is fulfilled
\begin{equation}
Q_{0} \geq\chi:=\left(  \frac{d-2}{2}\right)  ^{2} -\frac{1}{4}-\xi\left[
(d-1)(d-2)+\tilde{R}\right]  \, ,
\end{equation}
the frequencies are real for any effective mass satisfying the BF bound. If
$Q_{0}<\chi$, the positivity of $\omega^{2}$ compels the effective squared
mass to satisfy a more stringent bound. This is summarized by the bound
\begin{equation}
m_{\text{\textit{eff}}}^{2}>m_{BF}^{2}+m_{\Sigma,\xi}^{2} \,,
\label{mascotafastnon}%
\end{equation}
where
\begin{equation}
m_{\Sigma,\xi}^{2}l^{2}=\left\{
\begin{array}
[c]{cc}%
\left[  \sqrt{\left(  \frac{d-2}{2}\right)  ^{2}-Q_{0}-\xi\left[
(d-1)(d-2)+\tilde{R}\right]  }-\frac{1}{2}\right]  ^{2} & :Q_{0}<\chi\\
0 & :Q_{0}\geq\chi
\end{array}
\right.  \, . \label{msigmachi}%
\end{equation}

Stability is then guaranteed for the fast-fast fall-off, provided the bound
(\ref{mascotafastnon}), with (\ref{msigmachi}) is fulfilled.

\begin{itemize}
\item Slow-fast fall-off
\end{itemize}

\begin{figure}[ptb]
\includegraphics[scale=0.50]{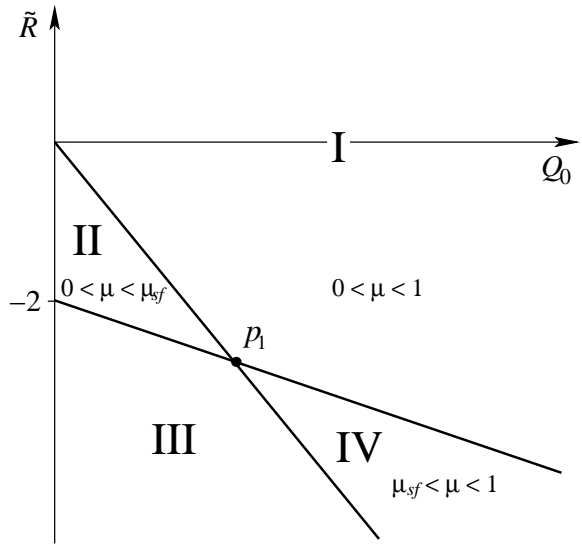}
\caption{Slow-fast behavior: Stability is guaranteed within regions I, II, and
IV for different bounds in the squared effective mass. In region III slow-fast
behavior is unstable.}%
\label{sfregions}%
\end{figure}

Slow fall-off for the scalar field at one side of the neck, and fast fall-off
at the other side, is admissible when the coupling constant $\xi$ is fixed as
in Eq. (\ref{beta}) and $0<\mu<1$. By virtue of Eq. (\ref{mueff}) this
corresponds to the following range of effective masses
\begin{equation}
m_{BF}^{2}l^{2}<m_{\text{\textit{eff}}}^{2}l^{2}<m_{BF}^{2}l^{2}+1\,.
\label{Upper-lower-bound}%
\end{equation}
The quantization relation reads
\begin{equation}
\nu=n\,,
\end{equation}
and leads to the following spectrum
\begin{equation}
\omega^{2}=\left(  n+\frac{1}{2}\right)  ^{2}-\left(  \frac{d-2}{2}\right)
^{2}+Q+\xi_{0}\left[  \left(  d-1\right)  \left(  d-2\right)  +\tilde
{R}\right]  \,. \label{rsfrec}%
\end{equation}
The frequency $\omega$ depends on the mass of the scalar field only through
$\xi_{0}$ (see Eqs. (\ref{beta}) and (\ref{lambdas})). The range of values of
$\mu$, for which frequencies are real, depends on the Ricci scalar (assumed to
be constant) and the lowest eigenvalue of the Laplace operator on the base
manifold. The half-plane spanned by $(Q_{0},\tilde{R})$ can be divided into
four regions delimited by the straight lines\footnote{These lines are obtained
demanding the vanishing of $\omega$ for the critical values $\mu=1$ and
$\mu=0$.}
\begin{align}
\tilde{R}  &  =-4\frac{d-2}{d-3}Q_{0}\,,\label{Line SF (0)}\\
\tilde{R}  &  =-4\frac{d}{d-1}Q_{0}-2\,, \label{Line SF (-2)}%
\end{align}
as it is depicted in Fig. \ref{sfregions}. The boundaries set by the lines
(\ref{Line SF (0)}) and (\ref{Line SF (-2)}) are included in the regions I and
III and the intersection point
\begin{equation}
p_{1}=\left(  \frac{(d-1)(d-3)}{4},-(d-1)(d-2)\right)  \,, \label{p1}%
\end{equation}
belongs to region I.

In regions II and IV, using
\begin{equation}
\mu_{sf}:=\frac{1}{2}\left[  d-\frac{\tilde{R}+(d-1)(d-2)}{\tilde{R}%
+4Q_{0}+d-1}\right]  \,,
\end{equation}
we have more stringent bounds than (\ref{Upper-lower-bound}) and in region
III, scalar fields with slow-fast fall-off are unstable. The ranges of $\mu$
for stable slow-fast scalar fields in the different regions are summarized in
table I.

\begin{table}[ptb]%
\begin{equation}%
\begin{tabular}
[c]{|c|c|}\hline
Region & Range of stability\\\hline
I & $0<\mu<1$\\\hline
II & $0<\mu<\mu_{sf}$\\\hline
III & -\\\hline
IV & $\mu_{sf}<\mu<1$\\\hline
\end{tabular}
\ \ \ \nonumber\label{tabla1}%
\end{equation}
\newline\vspace{-0.65cm}\caption{Slow-fast behavior: Stability ranges for the
different regions of the half-plane $(Q_{0},\tilde{R})$}%
\label{TableSF}%
\end{table}

\bigskip

Note that wormholes with base manifolds $\Sigma_{d-2}$ of
nonnegative scalar curvature, as it is for a torus $T^{d-2}$ or a
sphere $S^{d-2}$ of arbitrary radius  (which are not solutions of
the higher-dimensional gravitational theory considered), fall within
region I. Therefore, slow-fast scalar field perturbations are stable
regardless the size of the neck, $\rho_{0}$.

Notice that hyperbolic spaces $H_{d-2}$ of radius $r_{0}$, generate
the same line as in Eq. (\ref{Line SF (0)}) in the parameter space
$(Q_{0},\tilde{R})$. Thus, the stability of the slow-fast fall-off
excitation depends on $r_{0}$. Indeed, if $\Sigma_{d-2}$ is a
hyperbolic space of radius $r_{0}^{2}\geq\frac{d-3}{d-1}$, the
slow-fast excitation on this gravitational soliton is stable.
Quotients of $H_{d-2}$ giving to a closed surface of finite volume,
and then making (\ref{metricworm}) to describe a wormhole, lie in
axis $Q_{0}=0$. For those quotients, stability of the slow-fast
excitation depends on the neck radius: if
$\rho_{0}>l\sqrt{\frac{(d-2)(d-3)}{2}}$ the wormhole lies in region
II; otherwise, it lies in region III.

Base manifolds of the form $S^{1}\times H_{d-3}$, with $H_{d-3}$ of radius
$r_{0}$, are characterized by the line $\tilde{R}=-4\frac{d-3}{d-4}Q_{0}$, so
that they fall within region II provided the radius fulfills $r_{0}^{2}%
>\frac{3}{2}\frac{d-4}{d-1}$; else, they belong to region III.

It is interesting to pay special attention to base manifolds that make the
 metric (\ref{metricworm}) to be a vacuum solution of a special
class of higher-dimensional gravity theories \cite{DOTWORM}. We summarize some
of them in Appendix B. A thorough analysis which captures the features
described above, can be performed for base manifolds that are all the possible
products of constant curvature spaces.

In five dimensions, $\Sigma_{3}$ can be locally $H_{3}$ of unit
radius or $S^{1}\times H_{2}$, where the radius of $H_{2}$ is
$\frac{1}{\sqrt{3}}$. In the first case,  when $H_{3}$ is
non-quotiented the solution  describes a soliton and $Q_{0}=1$.
Then, it lies just at the edge of region I, but lies in the region
III if a quotient is taken  such that the metric (\ref{metricworm})
describes a wormhole and $Q_{0}=0$. The solution with $S^{1}\times
H_{2}$ lies in any case within region III.

\begin{figure}[ptb]
\includegraphics[scale=1.2]{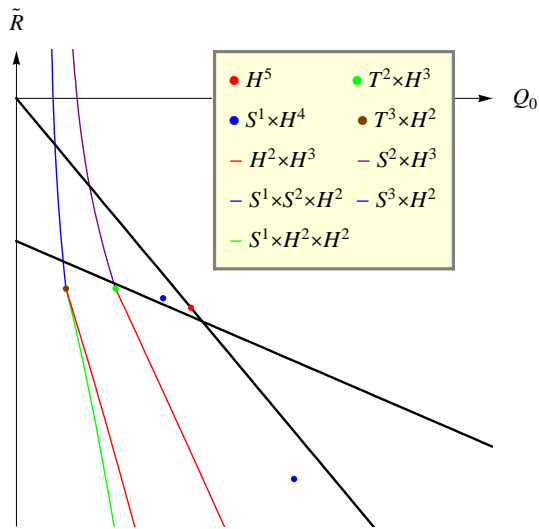}
\caption{Scalar fields with slow-fast behavior for seven-dimensional wormholes
in vacuum. The base manifolds are all the possible products of constant
curvature spaces.}%
\label{figd7sf}%
\end{figure}

In seven dimensions, there is a richer family of base manifolds
making the metric (\ref{metricworm}) a vacuum solution. They are not
just scattered points in the half-plane $(Q_{0},\tilde{R})$. There
are also one-parameter classes tracing curves in the half-plane
$(Q_{0},\tilde{R})$. For the solutions described in Fig.
\ref{figd7sf} (see also appendix B) we are considering the
hyperbolic factors as non-compact spaces, i.e. (\ref{metricworm})
corresponds to a gravitational soliton with an end of the world
brane. Then, each factor $H_{n}$ of radius $r_{H_{n}}$, contributes
with a term $\frac{1}{r_{H_{n}}^{2}}\left( \frac{n-1}{2}\right)
^{2}$ to $Q_{0}$. In order to obtain wormhole solutions one needs to
consider quotients that make $\Sigma_{5}$ a closed space of finite
volume. Then one would have $Q_{0}=0$ and the representation of the
corresponding wormhole solutions is the projection of the curves of
Fig. \ref{figd7sf} onto the vertical axis. In either case, for the
one-parameter families of base manifolds with sphere factors, one
can take the radius of the sphere to be sufficiently small so as to
lie in region II, or even within region I, where stability is
guaranteed for $0<\mu<1$.

\begin{itemize}
\item Slow-slow fall-off
\end{itemize}

The improved boundary conditions (\ref{flux2}) admit scalar fields with slow
fall-off at both sides of the wormhole when the coupling constant $\xi$ is
fixed as in Eq. (\ref{beta}) and for masses such that $0<\mu<1$. The spectrum
is obtained from the quantization relation
\begin{equation}
\mu+\nu=n\,,
\end{equation}
so that the frequencies read
\begin{equation}
\omega^{2}\!=\!\left(  \!n+\frac{1}{2}-\mu\right)  ^{2}\!\!-\left(  \frac
{d-2}{2}\right)  ^{2}\!\!+Q+\xi_{0}\left[  \left(  d-1\right)  \left(
d-2\right)  +\tilde{R}\right]  \!. \label{ssfrec}%
\end{equation}
Now, the frequency $\omega$ depends on the mass of the scalar field not only
through $\xi_{0}$ but also through the first term. Given $d$, $Q_{0}$ and
$\tilde{R}$, the stability of the scalar perturbation depends upon the sign of
the following cubic polynomial
\begin{equation}
P\left(  \mu\right)  :=-\mu^{3}+\frac{\left(  d+2\right)  }{2}\mu^{2}%
+\frac{\left(  1-3d-\tilde{R}-4Q_{0}\right)  }{4}\mu+\frac{\left(  d-1\right)
}{8}\tilde{R}+\frac{\left(  d-1+dQ_{0}\right)  }{4}\,, \label{P(mu)}%
\end{equation}
which can be monotonically decreasing or present two local extrema (a minimum
and a maximum). The ranges of the parameter $\mu$ in which $P(\mu)$ is
positive, divide the half-plane spanned by $(Q_{0},\tilde{R})$ in five
regions, which are {delimited} by the straight lines (\ref{Line SF (0)}) and
(\ref{Line SF (-2)}), and a nearly straight curve that is obtained from
$P(\mu_{-})=0$, where $\mu_{-}$ is the position of the minimum of
(\ref{P(mu)}):
\begin{equation}
\mu=\mu_{-}:=\frac{1}{6}\left(  d+2-\sqrt{d^{2}-5d+7-3\tilde{R}-12Q_{0}%
}\right)  \,.
\end{equation}

\begin{figure}[ptb]
\includegraphics[scale=0.50]{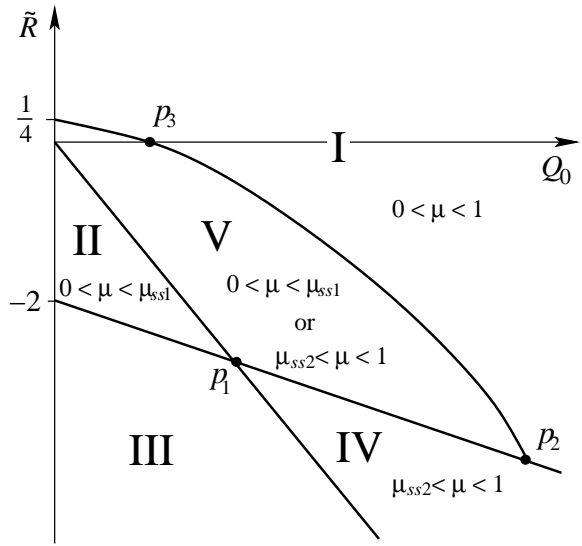}
\caption{The parameter space for slow-slow behavior: Stability is guaranteed
within regions I, II, IV, and V for different bounds in the squared effective
mass, respectively. Slow-slow behavior is unstable in region III. The points
$p_{1}$ and $p_{2}$ are defined in Eqs. (\ref{p1}), and (\ref{pedos}),
respectively. The point $p_{3}$ lies on the horizontal axis with
$Q_{0}<((d-3)/4)$.}%
\label{ssregions}%
\end{figure}

As it is depicted in Fig. \ref{ssregions}, this curve intersects the vertical
axis at $\tilde{R}=\frac{1}{4}$, and the straight line (\ref{Line SF (-2)}) at
the point
\begin{equation}
p_{2}=\left(  \frac{3}{4}(d-1)^{2},-2-3d(d-1)\right)  \ , \label{pedos}%
\end{equation}
which always lies below the point $p_{1}$, defined in Eq. (\ref{p1}) where the
lines (\ref{Line SF (0)}) and (\ref{Line SF (-2)}) intersect.

The detailed analysis of cases, although simple, is a bit clumsy. Let us just
summarize the results for the different regions. In region I, including its
boundary, the slow-slow excitation is stable for all $\mu$ satisfying the
bound (\ref{Upper-lower-bound}). For region II, stability is achieved for
excitations with a more stringent upper bound than in (\ref{Upper-lower-bound}%
), given by
\begin{equation}
0<\mu<\mu_{ss1}\,, \label{SS-bound-Region II}%
\end{equation}
where $\mu_{ss1}$ corresponds to the smallest root of the cubic polynomial
$P(\mu)$ (\ref{P(mu)})\footnote{The polynomial $P\left(  \mu\right)  $ admits
three roots $\mu_{ss1}$, $\mu_{ss2}$ and $\mu_{ss3}$. The largest root is real
and fulfills $\mu_{ss3}>1$.}, which in this region satisfies $\mu_{ss1}<1$.
The piece of the straight line (\ref{Line SF (0)}) that joins the origin with
the point $p_{1}$ is included within this region, including the origin but not
the point $p_{1}$.

In region IV, the stable excitations satisfy a more stringent lower bound than
in (\ref{Upper-lower-bound}), which reads
\begin{equation}
\mu_{ss2}<\mu<1\,, \label{SS-bound-Region IV}%
\end{equation}
where $\mu_{ss2}$ is the second root of the cubic polynomial $P(\mu)$. In this
case $\mu_{ss2}>0$. The segment of the straight line (\ref{Line SF (-2)}) is
included in this region, provided $\frac{\left(  d-1\right)  \left(
d-3\right)  }{4}<Q_{0}<\frac{3\left(  d-1\right)  ^{2}}{4}$.

In Region V, slow-slow excitations are stable provided the squared effective
mass lies in a range such that
\begin{equation}
0<\mu<\mu_{ss1}\ \ \text{\textrm{or\ } }\mu_{ss2}<\mu<1\,, \label{SSbV1}%
\end{equation}
where the bounds automatically fulfill $0<\mu_{ss1}<\mu_{ss2}<1$. The vertical
line $Q_{0}=0$ is included in this region, for $0<\tilde{R}<1/4$.

In region III, which includes its boundary, the scalar fields with slow-slow
behavior are unstable.

\begin{table}[ptb]%
\begin{equation}%
\begin{tabular}
[c]{|c|c|}\hline
Region & Range of stability\\\hline
I & $0<\mu<1$\\\hline
II & $0<\mu<\mu_{ss1}$\\\hline
III & -\\\hline
IV & $\mu_{ss2}<\mu<1$\\\hline
V & $0<\mu<\mu_{ss1}$ or $\mu_{ss2}<\mu<1$\\\hline
\end{tabular}
\ \ \ \nonumber\label{tabla}%
\end{equation}
\newline\vspace{-0.65cm}\caption{Slow-slow fall-off: Stability ranges for the
different regions of the half-plane $(Q_{0},\tilde{R})$}%
\label{Tabless}%
\end{table}

\bigskip

 As before, let us first consider some examples of generic base
manifolds, for which (\ref{metricworm}) is not necessarily a
solution of the higher-dimensional theory of gravity considered (see
appendix B). Note that spherically symmetric wormholes fall within
region I provided the radius of the neck is $\rho_{0}<2l$, otherwise
they belong to region V. If the base manifold is a torus $T^{d-2}$
the wormhole falls within region II, with $\mu_{ss1}=\frac{1}{2}$.
If $\Sigma_{d-2}$ is a  non-compact hyperbolic space of radius
$r_{0}^{2}>\frac{d-3}{d-1}$, the metric lies in region II; else, it
lies within region III and then scalar field perturbations with
slow-slow fall-off are unstable. If the base manifold is a quotient
of $H_{d-2}$ that includes the zero mode in the spectrum, for
$\rho_{0}>l\sqrt{\frac{(d-2)(d-3)}{2}}$ the wormhole one obtains
belongs to region II; otherwise it is located in region III.

Choosing the base manifold as $S^{1}\times H_{d-3}$, with $H_{d-3}$ of
arbitrary radius $r_{0}$, generates the line given by $\tilde{R}=-4\frac
{d-3}{d-4}Q_{0}$, which falls within region II provided the radius fulfills
$r_{0}^{2}>\frac{3}{2}\frac{d-4}{d-1}$; else, it belongs to region III.

Let us consider now base manifolds given by all the possible
products of constant curvature spaces, making the metric
(\ref{metricworm}) to be a vacuum solution of a special class of
higher-dimensional gravity theories \cite{DOTWORM}, for five and
seven dimensions.

In five dimensions, $\Sigma_{3}$ can be either locally $H_{3}$ of
unit radius or $S^{1}\times H_{2}$, where the radius of $H_{2}$ is
$\frac{1}{\sqrt{3}}$. In the first case the solution lies just at
the edge of region II for a non-quotiented $H_{3}$, but it would be
located in the region III if a suitable quotient  of $H_{3}$ makes
$\Sigma_{3}$ compact, so that the solution (\ref{metricworm})
describes a wormhole ($Q_{0}=0$ in that case). The case of
$\Sigma_{3}=S^{1}\times H_{2}$ belongs to region III, regardless
$H_{2}$ is quotiented or not,  i.e., regardless solution
(\ref{metricworm}) describes a gravitational soliton or a wormhole.

\begin{figure}[ptb]
\includegraphics[scale=1.2]{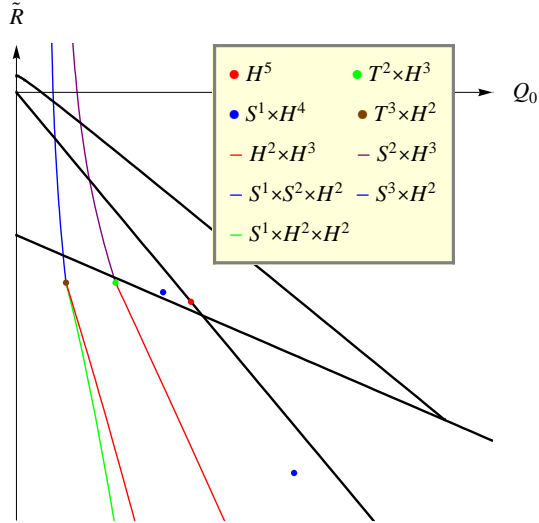}
\caption{Scalar fields with slow-slow behavior for seven-dimensional wormholes
in vacuum. The base manifolds are all the possible products of constant
curvature spaces.}%
\label{figd7ss}%
\end{figure}

In seven dimensions there are more possibilities. It is worth to remark that
the base manifold can possess a spherical factor, whose radius becomes a
modulus parameter. The radius of the sphere can be continuously shrunk to go
from region III to I, passing through regions II and V, as it is depicted in
Fig. \ref{figd7ss}.

\bigskip In summary, the spectrum of a free nonminimally coupled scalar field
has been obtained analytically. Three possible quantizations can be obtained
depending on the fall-off of the scalar field at both sides of the wormhole.
The stability of these scalar field perturbations on the wormhole is
guaranteed by requiring $\omega^{2}$ to be nonnegative, which imposes a bound
on the squared mass that depends on the geometry of the base manifold.

\section{Discussion}

The stability of scalar field perturbations on the class of wormholes
described by (\ref{metricworm}) was thoroughly analyzed. These were shown to
be stable provided the squared mass satisfies certain bounds, which
generically depend on the base manifold. The solutions to the corresponding
scalar field equations present two distinctive asymptotic behaviors at the
boundaries of the wormhole. These asymptotic behaviors were chosen so that the
energy flux vanishes at infinity, to ensure that we were dealing with
conserved energy excitations. Requiring the scalar field to vanish at infinity
would lead to the same modes for nonnegative (effective) squared masses.
However, for the range $m_{BF}^{2}<m^{2}<0$, the scalar field identically
vanishes at infinity, so that a Dirichlet condition would give no information
about the modes we found.

Minimally coupled free scalar perturbations, with masses satisfying the BF
bound $m_{BF}^{2}<m^{2}$, fulfill the boundary condition, but only when the
fast-fast fall-off behavior is selected. The stability of these perturbations
is guaranteed provided the mass is bounded as in Eqs. (\ref{imporvedbf}).
Similarly, nonminimally coupled scalar perturbations with fast-fast fall-off
are consistent with the vanishing of the energy flux at infinity for the full
range $m_{BF}^{2}<m_{eff}^{2}$. Within this range, stability is guaranteed
provided the mass is also bounded as in Eqs. (\ref{mascotafastnon}).

In the range $m_{BF}^{2}<m_{eff}^{2}<m_{BF}^{2}+\tfrac{1}{l^{2}}$, the
vanishing of the energy flux at infinity also admits nonminimally coupled
scalar perturbations with slow fall-off. Thus, three different quantizations
can be carried out for the scalar field, which are characterized by the
fall-off of the field, which can be fast or slow with respect to each
asymptotic region. The stability of these perturbations could set more
stringent upper and lower bounds for the range of squared effective masses, as
explained in Section \ref{nonminimal coupling}. These bounds depend on the
Ricci scalar $\tilde{R}$ and the lowest eigenvalue of the Laplace operator on
the base manifold $Q_{0}$, which characterize different wormholes. Then, the
half-plane spanned by $(Q_{0},\tilde{R})$ can be divided into regions,
according to ranges of squared effective masses for which the perturbations
are stable. This half-plane is divided into four regions for slow-fast
fall-off, and into five regions for slow-slow fall-off, as depicted in Figs.
\ref{sfregions} and \ref{ssregions}, respectively. The corresponding ranges
for the effective mass are given in Tables \ref{TableSF} and \ref{Tabless}.

For the range of masses admitting both asymptotic behaviors, the space of
physically admissible solutions is enlarged, and we found new interesting
configurations of scalar field excitations with conserved energy. This is
possible since the wormhole is asymptotically AdS at each side of the neck.
Consistency of scalar fields with slow fall-off was performed here through the
introduction of a nonminimal coupling with the scalar curvature. Nevertheless, it
does not escape to us that this could also be achieved for minimally coupled
scalar fields provided the energy flux is suitably regularized as in Ref
\cite{HMTZD}.

As it occurs for asymptotically AdS spacetimes
\cite{HMTZ2+1,HMTZlog,Hertog-Maeda,HMTZD,Amsel-Marolf}, it is natural to
expect that our results can be extrapolated to scalar fields with a
selfinteraction that can be unbounded from below. In this case, it would be
interesting to explore the subtleties due to the presence of a nontrivial
potential, since the asymptotic form of the scalar field obtained through the
linear equations could no longer be reliable. Indeed, for certain critical
values of the mass, the nonlinear terms in the potential could become relevant
in the asymptotic region, such that the scalar field would be forced to
develop additional logarithmic branches \cite{HMTZD}. These effects should
also be sensitive to the spacetime dimension, and for certain critical values
of the mass, they would be particularly relevant in the sense of the dual
conformal field theory. Nonetheless, note that the existence of asymptotically
AdS wormholes raises some puzzles concerning the AdS/CFT correspondence
\cite{WY, MM, Polchinski}.

It would be very interesting to further investigate non-perturbative
instabilities of these AdS wormholes due to brane creation (See
e.g., \cite{MM}). This is at least expected for the wormholes whose
base manifold has a negative Ricci scalar. Although the AdS/CFT dual
description of those backgrounds is unknown, the corresponding CFT
would be generically defined on a negatively curved space, and
conformally coupled scalar fields on negatively curved spaces could
cause tachyonic instabilities.

\bigskip

If the base manifold of the wormhole (\ref{metricworm}) is
restricted such that the metric solves the field equations in vacuum
for a special class of higher-dimensional gravity theories in odd
dimensions \cite{DOTWORM}, then the scalar excitations with
fast-fast fall-off are shown to be stable, provided the mass
fulfills the bounds (\ref{imporvedbf}) and (\ref{mascotafastnon})
for minimal and nonminimal coupling, respectively. For example, an
exact solution is obtained if the base manifold is chosen as
$\Sigma_{d-2} =S^{1}\times H_{d-3}/\Gamma$, with $H_{d-3}$ of radius
$\left( d-2\right)^{-1/2}$. In this case only scalar field
perturbations with fast-fast fall-off are stable on the wormhole. As
explained above, slow fall-off scalar fields are stable for certain
range of squared masses for base manifolds that do not fall within
Region III of the half-plane spanned by $(Q_{0},\tilde{R})$ (See
Figs. \ref{sfregions} and \ref{ssregions}). For instance, another
exact solution is obtained for $\Sigma_{d-2}=H_{d-2}$ with unit
radius. For this spacetime with a single conformal boundary scalar
fields with slow-fast behavior are stable for the range of masses
that corresponds to Region I, i.e. $0<\mu<1$. Slow-slow fall-off
excitations are stable for $0<\mu<\mu_{ss1}$, the range
corresponding to Region II. If $H_{d-2}$ is quotiented to obtain a
smooth closed surface with finite volume,  such that
(\ref{metricworm}) describes a wormhole, all scalar excitations with
slow fall-off are unstable, since the corresponding wormhole falls
within Region III.

As explained in Appendix B, there is a wide family of base manifolds
making the metric (\ref{metricworm}) to be a solution in vacuum. As
an example, in the seven-dimensional case, base manifolds given by
all the possible products of constant curvature spaces were
analyzed. It is possible to find another solution whose base
manifold is of the form $S^{1}\times H_{4}$, but where the
hyperbolic space is of unit radius. In this case, for noncompact
$H_{4}$, one obtains a soliton on which scalar field perturbations
with slow fall-off are stable in the range corresponding to Region
II. If the hyperbolic space is quotiented to obtain a smooth closed
surface with finite volume, then the corresponding wormhole falls
within Region III. It can also be seen that the base manifold admits
two- or three-spheres as a factor. In this case the radius of the
sphere is a modulus parameter that can be continuously shrunk so as
to move along different regions of the $(Q_{0},\tilde{R})$
half-plane. Thus, for sufficiently large spheres, scalar fields with
slow fall-off are unstable; nonetheless, their radius can be shrunk
so as the wormhole reaches Region II. The radius of the sphere can
further be shrunk to go to Region V for slow-slow behavior, as well
as to reach Region I for slow-fast and slow-slow fall-off, where the
scalar excitations are stable for all $0<\mu<1$.

It is also natural to wonder about the stability of the wormhole against
gravitational perturbations. One might be worried because in some of regions
of the $(Q_{0},\tilde{R})$ half-plane, the range of masses of stable
excitations is smaller than the range of masses of satisfying the boundary
conditions. More precisely, for $\mu_{ss1}<\mu<1$ in region II, for $0<\mu
<\mu_{ss2}$ in region IV and for $\mu_{ss1}<\mu<\mu_{ss1}$ in region V there
exist conserved energy excitations modes with $\omega^{2}<0$. However, it is
not at all obvious if they could be responsible for a gravitational
instability. For the class of theories under consideration, the degrees of
freedom of the graviton could depend on the background geometry (see e.g.
\cite{DEGENERATEDDYNAMICAL}, \cite{Olivera}), so that the dynamics of the
perturbations has to be analyzed from scratch. Moreover, if the dynamics of
some scalar perturbations of the wormhole solutions were reduced to the scalar
field equations we considered, typically this would be so for precise values
of the scalar masses. Therefore, for wormholes lying in regions II, IV and V,
only after knowing those precise masses one could say something about the
stability against these specific modes.

One could wonder about the chances of the wormholes being supersymmetric. It
is simple to check that the wormhole solves the field equations of the
corresponding locally supersymmetric extension in five \cite{CHAM2} and higher
odd dimensions \cite{SEVENANDELEVEN, ALLODD}. If the wormhole had some
unbroken supersymmetries, its stability would be guaranteed nonperturbatively.
However, a quick analysis shows that the wormhole in vacuum breaks all the
supersymmetries. Nonetheless, one cannot discard that supersymmetry could be
restored by switching on the torsion as in Ref. \cite{Paralelizable}, or by
considering nontrivial gauge fields without backreaction \cite{Olivera2},
\cite{Camell}.

\bigskip

It would also be interesting to explore whether stability holds along the
lines discussed here for a different class of wormholes in vacuum which has
been recently found \cite{DOT2}. For pure Gauss-Bonnet gravity, it has also
been shown that wormhole solutions with a jump in the extrinsic curvature
along a \textquotedblleft thin shell of nothingness\textquotedblright\ exist
\cite{GW}, and this has also been extended to the full Einstein-Gauss-Bonnet
theory in five dimensions \cite{GGGW}. For this theory, it is possible to have
wormholes made of thin shells of matter fulfilling the standard energy
conditions \cite{Simeone1, Simeone2}. For smooth matter distributions,
wormholes that do not violate the weak energy condition also exist\footnote{It
has been recently shown that this could also hold in four-dimensional
conformal gravity \cite{LOBOWEYL}. Wormhole solutions in higher dimensions
have also been discussed in the context of braneworlds, see e.g.,
\cite{LoboBW} and references therein.}, provided the Gauss-Bonnet coupling
constant is negative and bounded according to the shape of the solution
\cite{Bhawal-Kar, HIDEKI2}. Exact wormhole solutions in vacuum can also be
obtained for the Einstein-Gauss-Bonnet theory in higher dimensions
\cite{DOThd}, provided the Gauss-Bonnet coupling is chosen such that the
theory has a unique AdS vacuum, as in Ref. \cite{BHS}; and in turn, it has
been recently proved that this choice is a necessary condition for them to
exist \cite{HIDEKI2}.

\bigskip

\textit{Acknowledgments.-- }We thank  Hideki Maeda, Don Marolf and
David Tempo for helpful remarks.  We also thank the referee of this
work for very helpful comments. D.H.C. is  funded by the Seventh
Framework Programme under grant agreement number
PIEF-GA-2008-220702. J.O. thanks the support of projects MECESUP
UCO-0209 and MECESUP USA-0108. This work was partially funded by
FONDECYT grants 1051056, 1061291, 1071125, 1085322, 3060009,
3085043; Secyt-UNC and CONICET. The Centro de Estudios
Cient\'{\i}ficos (CECS) is funded by the Chilean Government through
the Millennium Science Initiative and the Centers of Excellence Base
Financing Program of CONICYT. CECS is also supported by a group of
private companies which at present includes Antofagasta Minerals,
Arauco, Empresas CMPC, Indura, Naviera Ultragas and Telef\'{o}nica
del Sur. CIN is funded by CONICYT and the Gobierno Regional de Los
R\'{\i}os.

\appendix

\section{Asymptotic Expansions for the Generalized Legendre functions}

\label{A}

The general solution of the Legendre equation (\ref{legendre}) is given by a
linear combination of the associated Legendre functions of first and second
kind, $P_{\nu}^{\mu}(x)$ and $Q_{\nu}^{\mu}(x)$, respectively. When the
positive parameter $\mu$ is not an integer, for our purposes it is convenient
to write the general solution as a linear combination of $P_{\nu}^{\mu}(x)$
and $P_{\nu}^{-\mu}(x)$, where
\begin{equation}
P_{\nu}^{-\mu}(x)=\frac{\Gamma(\nu-\mu+1)}{\Gamma(\nu+\mu+1)}\left(  \cos
(\mu\pi)P_{\nu}^{\mu}(x)-\frac{2}{\pi}\sin(\mu\pi)Q_{\nu}^{\mu}(x)\right)  \,.
\end{equation}
The asymptotic behavior of $P_{\nu}^{\mu}(x)$ as $x$ goes to $+1$ is,
\begin{equation}
P_{\nu}^{\mu}(x)\underset{x\rightarrow+1}{\sim}(1-x)^{-\mu/2}\left(  A_{0}%
(\mu)+A_{1}(\mu,\nu)(1-x)+\mathcal{O}[(1-x)^{2}]\right)  \,, \label{As1}%
\end{equation}
with
\begin{align}
A_{0}(\mu)  &  =\frac{2^{\mu/2}}{\Gamma(1-\mu)}\,,\label{A0}\\
A_{1}(\mu,\nu)  &  =-\frac{\mu\ 2^{\mu/2-2}}{\Gamma(1-\mu)}-\frac{\nu
(\nu+1)\ 2^{\mu/2-1}}{\Gamma(2-\mu)}\,. \label{A1}%
\end{align}
On the other hand, for $x\rightarrow-1$,
\begin{align}
P_{\nu}^{\mu}(x)\underset{x\rightarrow-1}{\sim}  &  (1+x)^{-\mu/2}\left(
B_{0}(\mu,\nu)+B_{1}(\mu,\nu)(1+x)+\mathcal{O}[(1+x)^{2}]\right) \nonumber\\
&  +(1+x)^{\mu/2}\left(  D_{0}(\mu,\nu)+D_{1}(\mu,\nu)(1+x)+\mathcal{O}%
[(1+x)^{2}]\right)  \ , \label{as3}%
\end{align}
where
\begin{align}
B_{0}(\mu,\nu)  &  =\frac{\pi2^{\mu/2}}{\sin(\pi\mu)\Gamma(-\nu)\Gamma
(1+\nu)\Gamma(1-\mu)}\ ,\label{B0}\\
B_{1}(\mu,\nu)  &  =\frac{\pi}{\sin(\pi\mu)\Gamma(-\nu)\Gamma(\nu+1)}\left(
\frac{\mu\ 2^{\mu/2-2}}{\Gamma(1-\mu)}+\frac{(\mu+\nu)(\mu-\nu-1)\ 2^{\mu
/2-1}}{\Gamma(2-\mu)}\right)  \ ,\label{B1}\\
D_{0}(\mu,\nu)  &  =-\frac{\pi2^{-\mu/2}}{\sin(\pi\mu)\Gamma(-\mu-\nu
)\Gamma(1+\nu-\mu)\Gamma(1+\mu)}\ ,\label{D0}\\
D_{1}(\mu,\nu)  &  =-\frac{\pi}{\sin(\pi\mu)\Gamma(-\mu-\nu)\Gamma(1+\nu-\mu
)}\left(  \frac{\mu\ 2^{-\mu/2-2}}{\Gamma(1+\mu)}-\frac{\nu(\nu+1)\ 2^{-\mu
/2-1}}{\Gamma(2+\mu)}\right)  \ . \label{D1}%
\end{align}
In the case of $\mu=k$, with $k$ a positive integer, we write the general
solution in terms of $P_{\nu}^{k}(x)$ and $Q_{\nu}^{k}(x)$. For $x\rightarrow
+1$ the asymptotic behavior reads
\begin{align}
P_{\nu}^{k}(x)\underset{x\rightarrow+1}{\sim}  &  (1-x)^{k/2}\left(
E_{0}(k,\nu)+\mathcal{O}[(1-x)]\right)  \,,\label{as5}\\
Q_{\nu}^{k}(x)\underset{x\rightarrow+1}{\sim}  &  (1-x)^{-k/2}\left(
F_{0}(k,\nu)+\mathcal{O}[(1-x)]\right) \nonumber\\
&  +\log(1-x)(1-x)^{k/2}\left(  G_{0}(k,\nu)+\mathcal{O}[(1-x)]\right)  \,,
\label{as6}%
\end{align}
where
\begin{align}
E_{0}(k,\nu)  &  =\frac{(\nu+k)\cdots(\nu+1-k)}{2^{k/2}k!}\,,\label{E0}\\
F_{0}(k,\nu)  &  =(-1)^{k}2^{k/2-1}(k-1)!\,,\label{F0}\\
G_{0}(k,\nu)  &  =\frac{(\nu+k)\cdots(\nu+1-k)}{2^{k/2+1}k!}\,. \label{G0}%
\end{align}
Similarly, for $x\rightarrow-1$, for our analysis one only needs the
asymptotic behavior of $P_{\nu}^{k}(x)$, given by
\begin{align}
P_{\nu}^{k}(x)\underset{x\rightarrow-1}{\sim}  &  (1+x)^{-k/2}\left(
H_{0}(k,\nu)+\mathcal{O}[(1+x)]\right) \nonumber\\
&  +\log(1+x)(1+x)^{k/2}\left(  K_{0}(k,\nu)+\mathcal{O}[(1+x)]\right)  \,,
\label{as7}%
\end{align}
where
\begin{align}
H_{0}(k,\nu)  &  =\frac{(-1)^{k}2^{k/2}(k-1)!(\nu+k)\cdots(\nu+1-k)}%
{\Gamma(k-\nu)\Gamma(k+\nu+1)}\ ,\\
K_{0}(k,\nu)  &  =\frac{(-1)^{k}(\nu+k)\cdots(\nu+1-k)}{{2^{k/2}k!\Gamma
(-\nu)\Gamma(\nu+1)}}\,. \label{K0}%
\end{align}

\section{Wormhole solutions in vacuum and their stability}

\label{B}

In this appendix we briefly summarize the wormhole solution in vacuum found in
Ref. \cite{DOTWORM}. The wormhole metric reads
\begin{equation}
ds^{2}=l^{2}\left[  -\cosh^{2}\left(  \rho-\rho_{0}\right)  dt^{2}+d\rho
^{2}+\cosh^{2}\left(  \rho\right)  d\Sigma_{2n-1}^{2}\right]
\ ,\label{metricwormdrho0}%
\end{equation}
where $\rho_{0}$ is an integration constant and $d\Sigma_{2n-1}^{2}$ stands
for the line element of the base manifold. This is an exact solution for a
very special class of gravity theories among the Lovelock family in higher odd
dimensions $d=2n+1$. The relative couplings of the Lovelock series are chosen
so that the action has the highest possible power in the curvature and
possesses a unique AdS vacuum of radius $l$. The apparent mass at each side of
the wormhole vanishes for $\rho_{0}=0$ and the metric reduces to
(\ref{metricworm}) which acquires reflection symmetry. The metric of the base
manifold must solve the following scalar equation
\begin{equation}
\epsilon_{m_{1}\cdot\cdot\cdot m_{2n-1}}\bar{R}^{m_{1}m_{2}}\cdot\cdot
\cdot\bar{R}^{m_{2n-3}m_{2n-2}}\tilde{e}^{m_{2n-1}}=0\,.\label{TrCS}%
\end{equation}
Here $\bar{R}^{mn}:=\tilde{R}^{mn}+\tilde{e}^{m}\tilde{e}^{n}$, where
$\tilde{R}^{mn}$ and $\tilde{e}^{m}$ are the curvature two-form and the
vielbein of $\Sigma_{_{2n-1}}$, respectively. This equation admits a wide
class of solutions, and it is simple to verify that $\Sigma_{2n-1}=H_{2n-1}$
and $\Sigma_{2n-1}=S^{1}\times H_{2n-2}$ solve (\ref{TrCS}) provided the radii
of the hyperbolic spaces $H_{2n-1}$ and $H_{2n-2}$ are given by $r_{H_{2n-1}%
}=1$ and $r_{H_{2n-2}}=(2n-1)^{-1/2}$, respectively \footnote{Note
that, as explained in \cite{DOTWORM}, the field equations acquire
certain class of degeneracy around the solution with
$\Sigma_{2n-1}=H_{2n-1}$.}. The hyperbolic factors of the base
manifold must be quotiented in order
(\ref{metricwormdrho0}) to describe a wormhole, otherwise the spacetime%
 would correspond to a gravitational soliton possessing a single
conformal boundary. In this appendix we present the solutions of Eq.
(\ref{TrCS}) for base manifolds that are all the possible products
of constant curvature spaces in five and seven dimensions.

\bigskip

In five dimensions, Eq. (\ref{TrCS}) reduces to
\begin{equation}
\tilde{R}=-6\,, \label{ecu5}%
\end{equation}
where $\tilde{R}$ is the Ricci scalar of the three-dimensional base manifold
$\Sigma_{3}$. If the base manifold is a product of lower dimensional spaces of
constant curvature, then it is simple to verify that Eq. (\ref{ecu5}) is
solved only if $\Sigma_{3}=H_{3}$ with unit radius, or $\Sigma_{3}=S^{1}\times
H_{2}$ with $r_{H_{2}}=3^{-1/2}$.

In the case of $\Sigma_{3}=H_{3}$ of infinite volume, as shown in
Fig. \ref{figd7sf}, the soliton lies in region I, where scalar field
perturbations with slow-fast asymptotic behavior are stable for the
range (\ref{Upper-lower-bound}). In the case of the slow-slow
fall-off, the soliton belongs to region II, so that in order to
reach stability, the bound (\ref{SS-bound-Region II}) must be
satisfied. Considering a smooth closed quotient of $H_{3}$ with
finite volume makes the spacetime (\ref{metricworm}) a wormhole
which falls in region III, where scalar fields with slow fall-off
are unstable.

For the remaining possibility, $\Sigma_{3}=S^{1}\times H_{2}$,
regardless the compactness of the hyperbolic manifold $H_{2}$ the
corresponding soliton always falls in region III.

\bigskip

In seven dimensions, Eq. (\ref{TrCS}) reads%
\begin{equation}
\mathcal{E}+12\tilde{R}+120=0\ , \label{ecu7}%
\end{equation}
where $\mathcal{E}:=\tilde{R}^{2}-4\tilde{R}_{mn}\tilde{R}^{mn}+\tilde
{R}_{\ pq}^{mn}\tilde{R}_{\ mn}^{pq}$ is the Gauss-Bonnet combination. In this
case, there are more interesting possibilities among the possible products of
lower dimensional spaces of constant curvature.

Let $M_{n}$ be an $n$-dimensional manifold of constant curvature $\tilde
{R}_{\ kl}^{ij}=\lambda_{n}\left(  \delta_{k}^{i}\delta_{l}^{j}-\delta_{l}%
^{i}\delta_{k}^{j}\right)  $. The solution of Eq. (\ref{ecu7}) for $\Sigma
_{5}=M_{5}$ is given by $\lambda_{5}=-1$, so that $\Sigma_{5}$ is locally a
hyperbolic space $H_{5}$ of radius $r_{H_{5}}=1$. Taking $\Sigma_{5}$ to be
locally of the form $S^{1}\times M_{4}$, one obtains
\begin{equation}
\left(  \lambda_{4}+1\right)  \left(  \lambda_{4}+5\right)  =0\,,
\end{equation}
which means that $M_{4}$ is locally given by $H_{4}$ whose radius can be
either $r_{H_{4}}=1$ or $r_{H_{4}}=5^{-1/2}$.

For $\Sigma_{5}=M_{3}\times M_{2}$, Eq. (\ref{ecu7}) reduces to
\begin{equation}
\lambda_{3}\lambda_{2}+\lambda_{2}+3\lambda_{3}+5=0\,. \label{M3xM2constraint}%
\end{equation}
This last equation can be solved for $\lambda_{3}\neq-1$ and $\lambda_{2}%
\neq-3$, leading to
\begin{equation}
\lambda_{3}=-\frac{\lambda_{2}+5}{\lambda_{2}+3}\,, \label{M3N2l3intermsofl2}%
\end{equation}
which defines a one-parameter family of solutions. Thus, for $0<\lambda
_{2}<\infty$ the sign of $\lambda_{3}$ is negative, and hence $M_{2}$ and
$M_{3}$ are locally $S^{2}$ and $H_{3}$, respectively. For $\lambda_{2}=0$,
then $M_{2}=T^{2}$ and $M_{3}=H_{3}$, with $\lambda_{3}=-5/3$. For the range
$-3<\lambda_{2}<0$, one obtains that both $M_{2}$ and $M_{3}$ are locally
hyperbolic spaces. If $-5<\lambda_{2}<-3$, then $M_{2}=H_{2}$ and $M_{3}
=S^{3}$; and for $\lambda_{2}=-5$, $\lambda_{3}$ vanishes so that we have
$M_{2}=H_{2}$ and $M_{3}=T^{3}$. Finally for $-\infty<\lambda_{2}<-5$ we
obtain again that $M_{2}=H_{2}$ and $M_{3}=H_{3}$ locally, but for a different
range of the radii as compared with the previous case.

In sum, the allowed base manifolds of the form $\Sigma_{5}=M_{3}\times M_{2}$
are described by one-parameter families, and the different possibilities as
well as the relationship between their radii are shown in the first two
columns of Table III.

The remaining possibilities are of the form $\Sigma_{5}=S^{1}\times
M_{2}\times\hat{M}_{2}$, so that Eq. (\ref{ecu7}) now reads
\begin{equation}
\hat{\lambda}_{2}\lambda_{2}+3\left(  \hat{\lambda}_{2}+\lambda_{2}\right)
+15=0\ .
\end{equation}
This equation can be solved for $\lambda_{2},\ \hat{\lambda}_{2}\neq-3$, also
giving a one-parameter family of spaces. The relationship between the
curvatures of $M_{2}$ and $\hat{M}_{2}$ reads
\begin{equation}
\lambda_{2}=-3\left(  \frac{\hat{\lambda}_{2}+5}{\hat{\lambda}_{2}+3}\right)
\,.
\end{equation}
Hence, for $0<\hat{\lambda}_{2}<\infty$, the sign of $\lambda_{2}$ is negative
so that $M_{2}$ and $\hat{M}_{2}$ are locally $H_{2}$ and $S^{2}$,
respectively. If $\hat{\lambda}_{2}=0$, then $\lambda_{2}=-5$, so that
$\hat{M}_{2}=T^{2}$ and $M_{2}=H_{2}$, locally. For the range $-3<\hat
{\lambda}_{2}<0$ one obtains that both $M_{2}$ and $\hat{M}_{2}$ are locally
hyperbolic. These latter possibilities are also described by a one-parameter
family, and they are shown in Table III, altogether with the relationships
between their radii.

\begin{table}[ptb]%
\begin{equation}%
\begin{tabular}
[c]{|c|c|c|c|}\hline
$\Sigma_{5}$ & radii & slow-fast regions & slow-slow regions\\\hline
$H_{5}$ & $r_{H_{5}}^{2}=1$ & I & II\\\hline
$S^{1}\times H_{4}$ & $r_{H_{4}}^{2}=\left\{
\begin{array}
[c]{c}%
\frac{1}{5}\\
1
\end{array}
\right.  $ & $%
\begin{array}
[c]{c}%
\text{III}\\
\text{II}%
\end{array}
$ & $%
\begin{array}
[c]{c}%
\text{III}\\
\text{II}%
\end{array}
$\\\hline
$T^{2}\times H_{3}$ & $r_{H_{3}}^{2}=\frac{3}{5}$ & III & III\\\hline
$S^{2}\times H_{3}$ & $r_{H_{3}}^{2}=\frac{3r_{S^{2}}^{2}+1}{5r_{S^{2}}^{2}%
+1}$ & III, II, I & III, II, V, I\\\hline
$S^{3}\times H_{2}$ & $r_{H_{2}}^{2}=\frac{r_{S^{3}}^{2}+1}{5r_{S^{3}}^{2}+3}$
& III, II, I & III, II, V, I\\\hline
$H_{3}\times H_{2}$ & $r_{H_{3}}^{2}=\frac{3r_{H_{2}}^{2}-1}{5r_{H_{2}}^{2}%
-1}$ & III & III\\\hline
$T^{3}\times H_{2}$ & $r_{H_{2}}^{2}=\frac{1}{5}$ & III & III\\\hline
$S^{1}\times H_{2}\times\hat{H}_{2}$ & $r_{\hat{H}_{2}}^{2}=\frac{3r_{H_{2}%
}^{2}-1}{3\left(  5r_{H_{2}}^{2}-1\right)  }$ & III & III\\\hline
$S^{1}\times S^{2}\times H_{2}$ & $r_{H_{2}}^{2}=\frac{3r_{S^{2}}^{2}%
+1}{3\left(  5r_{S^{2}}^{2}+1\right)  }$ & III, II, I & III, II, V, I\\\hline
\end{tabular}
\ \ \ \ \nonumber
\end{equation}
\newline\vspace{-0.65cm}\caption{Seven-dimensional wormholes in vacuum:
Allowed base manifolds made of products of lower dimensional constant
curvature spaces. The relationship between their radii is shown, as well as
the slow-fast and slow-slow regions in which these solutions can be found.}%
\end{table}

In what follows we describe the stability of the seven-dimensional wormhole in
vacuum against scalar field perturbations with slow-fast or slow-slow
fall-off, for all the possible base manifolds given by products of constant
curvature spaces which are listed in Table III. As explained in Section
\ref{nonminimal coupling}, scalar field perturbations with slow asymptotic
behavior are stable for certain ranges of $\mu$, provided the base manifold
possesses a Ricci scalar ($\tilde{R}$) and a lowest eigenvalue of the Laplace
operator ($Q_{0}$) such that it is located outside region III of the
$(Q_{0},\tilde{R})$ half plane. For the class of base manifolds under
consideration, this is depicted in Figs. \ref{figd7sf} and \ref{figd7ss} for
slow-fast and slow-slow fall-off, respectively.

The case $\Sigma_{5}=H_{5}$ defines a point in the $(Q_{0},\tilde{R})$
half-plane with coordinates $\left(  4,-20\right)  $. For slow-fast behavior,
this point belongs to region I, and for slow-slow fall-off lies on region II.
For $\Sigma_{5}=S^{1}\times H_{4}$, the hyperbolic space can be of different
radii, leading to two different points in the $(Q_{0},\tilde{R})$ half-plane,
with coordinates $\left(  \frac{9}{4},-12\right)  $ and $\left(  \frac{45}%
{4},-60\right)  $. The first case is located in region II, and the latter in
region III.

Base manifolds $\Sigma_{5}$ of the form $M_{3}\times M_{2}$, define curves in
the $(Q_{0},\tilde{R})$ half-plane which can be parameterized in terms of
$\lambda_{2}$, leading to
\begin{align}
\tilde{R}  &  =2\left(  \frac{\lambda_{2}^{2}-15}{\lambda_{2}+3}\right)
\,,\label{M3N2Riccisubs}\\
Q_{0}  &  =\left\vert \frac{\lambda_{2}+5}{\lambda_{2}+3}\right\vert \bar
{Q}_{0}\left(  M_{3}\right)  +|\lambda_{2}|\bar{Q}_{0}\left(  M_{2}\right)
\,, \label{M3N2Q0subs}%
\end{align}
where $\bar{Q}_{0}\left(  M_{n}\right)  $ denotes the lowest eigenvalue of the
Laplace operator on $M_{n}$ of curvature normalized to $\pm1,\ 0$.
Analogously, for base manifolds of the form $\Sigma_{5}=S^{1}\times
M_{2}\times\hat{M}_{2}$ the curves are parameterized according to
\begin{align}
R  &  =2\left(  \frac{\lambda_{2}^{2}-15}{\lambda_{2}+3}\right)  \,,\\
Q_{0}  &  =\left\vert \lambda_{2}\right\vert \bar{Q}_{0}\left(  M_{2}\right)
+3\left\vert \frac{\lambda_{2}+5}{\lambda_{2}+3}\right\vert \bar{Q}_{0}\left(
\hat{M}_{2}\right)  \,.
\end{align}
The curves are shown in Figs. \ref{figd7sf} and \ref{figd7ss}. The red curves,
corresponds to $\Sigma_{5}=H_{2}\times H_{3}$ which lie within region III
independently of the radius $r_{H_{2}}$. The piece on the left is for
$0<r_{H_{2}}<\tfrac{1}{\sqrt{5}}$ and the piece on the right is for $\tfrac
{1}{\sqrt{3}}<r_{H_{2}}$. These two red curves end on the points $\left(
\frac{5}{4},-10\right)  $ and $\left(  \frac{5}{3},-10\right)  $ (brown and
green) also corresponding to $\Sigma_{5}=H_{2}\times T^{3}$ and $\Sigma
_{5}=T^{2}\times H_{3}$ respectively.

The purple curve corresponds to $\Sigma_{5}=S^{2} \times H_{3}$, where the
radius of the sphere can be continuously shrunk to go from region III to I.
For two-spheres of radius fulfilling $5+2\sqrt{7}\leq r_{S^{2}}^{2}<\infty$,
the slow branch is unstable (region III). For $\frac{\left(  5+\sqrt
{65}\right)  }{10}<r_{S^{2}}^{2}<5+2\sqrt{7}$ the slow branch is stable
provided $\mu$ satisfies the bounds that correspond to region II. If the
radius of the sphere fulfills $r_{S^{2}}^{2}\leq\frac{\left(  5+\sqrt
{65}\right)  }{10}$, the wormhole admits slow-fast behavior with bounds on
$\mu$ corresponding to region I. Slow-slow behavior is also allowed for
$r_{S^{2}}^{2}=\frac{\left(  5+\sqrt{65}\right)  }{10}$ (region II),
$r_{0}^{2}<r_{S^{2}}^{2}<\frac{\left(  5+\sqrt{65}\right)  }{10}$ (region V),
and $r_{S^{2}}^{2}\leq r_{0}^{2}$ (region I), with $r_{0}^{2}\simeq0.665$.

Note that the analysis changes for quotients of hyperbolic spaces with finite
volume. Since in those cases $Q_{0}=0$ and the curves are projected onto the
vertical axis. Then, slow branches for $\Sigma_{5}=H_{2} \times H_{3}$ or
$\Sigma_{5}=T^{2}\times H_{3}$ lie in region III. In the case of $\Sigma
_{5}=S^{2} \times H_{3}$ the radii that define the transition between the
different regions are given by $r_{S^{2}}^{2}=1/3,$ $1/\sqrt{15}$, and
$16/\left(  1+\sqrt{3937}\right)  $.

The green curve in Figs. \ref{figd7sf} and \ref{figd7ss}, also ending at the
brown point $\left(  \frac{5}{4},-10\right)  $, describes base manifolds
$\Sigma_{5}=H_{2}\times H_{2}\times S^{1}$ and lies completely into region
III. The blue curve describes alternatively $\Sigma_{5}=H_{2}\times
S^{2}\times S^{1}$ or $\Sigma_{5}=H_{2}\times S^{3}$ . In these cases, the
radii that define the transition between the different regions for the
two-sphere are given by $r_{S^{2}}^{2}=\frac{1}{26}\left(  11+\sqrt
{329}\right)  $, $\frac{1}{15}\left(  10+\sqrt{185}\right)  $ and $r_{S^{2}%
}^{2}=r_{V}^{2}$, with $r_{V}^{2}\simeq0.563$. For the three-sphere the
corresponding radii are $r_{S^{3}}^{2}=\frac{3}{26}\left(  11+\sqrt
{329}\right)  $, $\frac{1}{5}\left(  10+\sqrt{185}\right)  $ and $r_{S^{3}%
}^{2}=3r_{V}^{2}$. In the case of hyperbolic spaces of finite volume, the
transition radii correspond to $r_{S^{2}}^{2}=\frac{1}{3}$, $\frac{\sqrt{15}%
}{15}$ and $\frac{1}{246}\left(  \sqrt{3937}-1\right)  $, and $r_{S^{3}}%
^{2}=1$, $\frac{\sqrt{15}}{5}$ and $\frac{1}{82}\left(  \sqrt{3937}-1\right)
$.

Note that the family of base manifolds considered in this appendix never falls
within region IV.


\begin{thebibliography}{99}                                                                                               %


\bibitem {MTY}M.~S.~Morris, K.~S.~Thorne and U.~Yurtsever,
Phys.\ Rev.\ Lett.\ \textbf{61}, 1446 (1988)
; M.~S.~Morris and K.~S.~Thorne,
Am.\ J.\ Phys.\ \textbf{56}, 395 (1988).


\bibitem {Visser}M.~Visser, \textquotedblleft\textit{Lorentzian wormholes:
From Einstein to Hawking}", Woodbury, USA: AIP (1995).

\bibitem {LOVELOCK}D.~Lovelock,
J.\ Math.\ Phys.\ \textbf{12} (1971) 498.


\bibitem {DOTWORM}G.~Dotti, J.~Oliva and R.~Troncoso,
Phys.\ Rev.\ D \textbf{75}, 024002 (2007).


\bibitem {BF1}P.~Breitenlohner and D.~Z.~Freedman,
Phys.\ Lett.\ B \textbf{115}, 197 (1982).


\bibitem {BF2}P.~Breitenlohner and D.~Z.~Freedman,
Annals Phys.\ \textbf{144}, 249 (1982).


\bibitem {MEZTOW}L.~Mezincescu and P.~K.~Townsend,
Annals Phys.\ \textbf{160}, 406 (1985).


\bibitem {KR}A.~Karch and L.~Randall,
JHEP \textbf{0106}, 063 (2001).


\bibitem {Porrati}M.~Porrati,
Mod.\ Phys.\ Lett.\ A \textbf{18}, 1793 (2003).
, M.~Porrati,
JHEP \textbf{0204}, 058 (2002).


\bibitem {MM}J.~M.~Maldacena and L.~Maoz,
JHEP \textbf{0402}, 053 (2004).


\bibitem {HTZBrane}M.~Hassaine, R.~Troncoso and J.~Zanelli,
Phys.\ Lett.\ B \textbf{596}, 132 (2004).


\bibitem {GW}E.~Gravanis and S.~Willison,
Phys.\ Rev.\ D \textbf{75}, 084025 (2007).


\bibitem {GGGW}C.~Garraffo, G.~Giribet, E.~Gravanis and S.~Willison,
\textquotedblleft Gravitational solitons and $C^{0}$ vacuum metrics in
five-dimensional Lovelock gravity,\textquotedblright\ arXiv:0711.2992 [gr-qc].
To be published in J. Math. Phys.


\bibitem {Mann1}R.~B.~Mann,
Class.\ Quant.\ Grav.\ \textbf{14}, L109 (1997).


\bibitem {Mann2}R.~B.~Mann,
Nucl.\ Phys.\ B \textbf{516}, 357 (1998).


\bibitem {Vanzo}L.~Vanzo,
Phys.\ Rev.\ D \textbf{56}, 6475 (1997)


\bibitem {BLP}D.~R.~Brill, J.~Louko and P.~Peldan,
Phys.\ Rev.\ D \textbf{56}, 3600 (1997).


\bibitem {Birmingham}D.~Birmingham,
Class.\ Quant.\ Grav.\ \textbf{16}, 1197 (1999).


\bibitem {Cai-Soh}R.~G.~Cai and K.~S.~Soh,
Phys.\ Rev.\ D \textbf{59}, 044013 (1999).


\bibitem {Aros}R.~Aros, R.~Troncoso and J.~Zanelli,
Phys.\ Rev.\ D \textbf{63}, 084015 (2001).

\bibitem {QNM}R.~Aros, C.~Martinez, R.~Troncoso and J.~Zanelli,
Phys.\ Rev.\ D \textbf{67}, 044014 (2003).


\bibitem {BM1}D.~Birmingham and S.~Mokhtari,
Phys.\ Rev.\ D \textbf{74}, 084026 (2006).


\bibitem {BM2}D.~Birmingham and S.~Mokhtari,
Phys.\ Rev.\ D \textbf{76}, 124039 (2007).


\bibitem {Gibbons-Hartnoll}G.~Gibbons and S.~A.~Hartnoll,
Phys.\ Rev.\ D \textbf{66}, 064024 (2002).


\bibitem {Kodama-Ishibashi}A.~Ishibashi and H.~Kodama,
Prog.\ Theor.\ Phys.\ \textbf{110}, 901 (2003).


\bibitem {Kodama}H.~Kodama, ``Perturbations and Stability of
Higher-Dimensional Black Holes,'' arXiv:0712.2703 [hep-th].


\bibitem {Groundstate}R.~Aros, C.~Martinez, R.~Troncoso and J.~Zanelli,
JHEP \textbf{0205}, 020 (2002).

\bibitem {Mann3}J.~S.~F.~Chan and R.~B.~Mann,
Phys.\ Rev.\ D \textbf{55}, 7546 (1997).
, J.~S.~F.~Chan and R.~B.~Mann,
Phys.\ Rev.\ D \textbf{59}, 064025 (1999).
, B.~Wang, E.~Abdalla and R.~B.~Mann,
Phys.\ Rev.\ D \textbf{65}, 084006 (2002).


\bibitem {HMTZD}M.~Henneaux, C.~Martinez, R.~Troncoso and J.~Zanelli,
Annals Phys.\ \textbf{322}, 824 (2007).


\bibitem {HMTZ2+1}M.~Henneaux, C.~Martinez, R.~Troncoso and J.~Zanelli,
Phys.\ Rev.\ D \textbf{65}, 104007 (2002).


\bibitem {HMTZlog}M.~Henneaux, C.~Martinez, R.~Troncoso and J.~Zanelli,
Phys.\ Rev.\ D \textbf{70}, 044034 (2004).


\bibitem {Hertog-Maeda}T.~Hertog and K.~Maeda,
JHEP \textbf{0407}, 051 (2004).


\bibitem {Amsel-Marolf}A.~J.~Amsel and D.~Marolf,
Phys.\ Rev.\ D \textbf{74}, 064006 (2006) [Erratum-ibid.\ D \textbf{75},
029901 (2007)].


\bibitem {WY}E.~Witten and S.~T.~Yau,
Adv.\ Theor.\ Math.\ Phys.\ \textbf{3}, 1635 (1999).


\bibitem {Polchinski}N.~Arkani-Hamed, J.~Orgera and J.~Polchinski,
JHEP \textbf{0712}, 018 (2007).


\bibitem {DEGENERATEDDYNAMICAL}J.~Saavedra, R.~Troncoso and J.~Zanelli,
J.\ Math.\ Phys.\ \textbf{42}, 4383 (2001).


\bibitem {Olivera}O.~Miskovic, R.~Troncoso and J.~Zanelli,
Phys.\ Lett.\ B \textbf{615}, 277 (2005).


\bibitem {CHAM2}A.~H.~Chamseddine,
Nucl.\ Phys.\ B \textbf{346}, 213 (1990).

\bibitem {SEVENANDELEVEN}R.~Troncoso and J.~Zanelli,
Phys.\ Rev.\ D \textbf{58}, 101703 (1998)

\bibitem {ALLODD}R.~Troncoso and J.~Zanelli,
Int.\ J.\ Theor.\ Phys.\ \textbf{38}, 1181 (1999).


\bibitem {Paralelizable}F.~Canfora, A.~Giacomini and R.~Troncoso,
Phys.\ Rev.\ D \textbf{77}, 024002 (2008).


\bibitem {Olivera2}O.~Miskovic, R.~Troncoso and J.~Zanelli,
Phys.\ Lett.\ B \textbf{637}, 317 (2006).


\bibitem {Camell}F. Canfora, A. Giacomini, J. Oliva and R. Troncoso, work in progress.

\bibitem {DOT2}G.~Dotti, J.~Oliva and R.~Troncoso,
Phys.\ Rev.\ D \textbf{76}, 064038 (2007).


\bibitem {Simeone1}M.~Thibeault, C.~Simeone and E.~F.~Eiroa,
Gen.\ Rel.\ Grav.\ \textbf{38}, 1593 (2006).


\bibitem {Simeone2}M.~G.~Richarte and C.~Simeone,
Phys.\ Rev.\ D \textbf{76}, 087502 (2007).


\bibitem {LOBOWEYL}F.~S.~N.~Lobo, ``General class of wormhole geometries in
conformal Weyl gravity,'' arXiv:0801.4401 [gr-qc].


\bibitem {LoboBW}F.~S.~N.~Lobo,
Phys.\ Rev.\ D \textbf{75}, 064027 (2007).


\bibitem {Bhawal-Kar}B.~Bhawal and S.~Kar,
Phys.\ Rev.\ D \textbf{46}, 2464 (1992).


\bibitem {HIDEKI2}H.~Maeda and M.~Nozawa,
  Phys.\ Rev.\  D {\bf 78}, 024005 (2008).

\bibitem {DOThd}G. Dotti, J. Oliva and R. Troncoso, \textquotedblleft
Wormholes and black holes with nontrivial boundaries for the
Einstein-Gauss-Bonnet theory in vacuum,\textquotedblright\ Preprint: CECS-PHY-07/21. Proceedings of the ``Seventh Alexander Friedmann International Seminar on Gravitation and Cosmology", to be published.

\bibitem {BHS}J.~Crisostomo, R.~Troncoso and J.~Zanelli,
Phys.\ Rev.\ D \textbf{62}, 084013 (2000).

\end{thebibliography}
\end{document}